\documentclass[aps,prx,a4paper,superscriptaddress,twocolumn,showpacs,amsmath,amssymb,longbibliography]{revtex4-1}

\usepackage{graphicx,epsfig}
\usepackage[usenames]{color}

\usepackage[english]{babel}

\usepackage{csquotes}
\MakeOuterQuote{"}

\DeclareMathOperator*{\argmin}{argmin}

\allowdisplaybreaks

\begin{document}

\newcommand{\E}{\mathcal{E}}
\newcommand{\G}{\mathcal{G}}
\newcommand{\Lag}{\mathcal{L}}
\newcommand{\M}{\mathcal{M}}
\newcommand{\N}{\mathcal{N}}
\newcommand{\U}{\mathcal{U}}
\newcommand{\R}{\mathcal{R}}
\newcommand{\F}{\mathcal{F}}
\newcommand{\V}{\mathcal{V}}
\newcommand{\C}{\mathcal{C}}
\newcommand{\I}{\mathcal{I}}
\newcommand{\s}{\sigma}
\newcommand{\up}{\uparrow}
\newcommand{\dw}{\downarrow}
\newcommand{\h}{\hat{\mathcal{H}}}
\newcommand{\himp}{\hat{h}}
\newcommand{\g}{\mathcal{G}^{-1}_0}
\newcommand{\D}{\mathcal{D}}
\newcommand{\A}{\mathcal{A}}
\newcommand{\projs}{\hat{\mathcal{S}}_d}
\newcommand{\proj}{\hat{\mathcal{P}}_d}
\newcommand{\K}{\textbf{k}}
\newcommand{\Q}{\textbf{q}}
\newcommand{\T}{\tau_{\ast}}
\newcommand{\io}{i\omega_n}
\newcommand{\eps}{\varepsilon}
\newcommand{\+}{\dag}
\newcommand{\su}{\uparrow}
\newcommand{\giu}{\downarrow}
\newcommand{\0}[1]{\textbf{#1}}
\newcommand{\ca}{c^{\phantom{\dagger}}}
\newcommand{\cc}{c^\dagger}
\newcommand{\fa}{f^{\phantom{\dagger}}}
\newcommand{\fc}{f^\dagger}
\newcommand{\aaa}{a^{\phantom{\dagger}}}
\newcommand{\aac}{a^\dagger}
\newcommand{\bba}{b^{\phantom{\dagger}}}
\newcommand{\bbc}{b^\dagger}
\newcommand{\da}{\hat{d}^{\phantom{\dagger}}}
\newcommand{\dc}{\hat{d}^\dagger}
\newcommand{\ha}{h^{\phantom{\dagger}}}
\newcommand{\hc}{h^\dagger}
\newcommand{\be}{\begin{equation}}
\newcommand{\ee}{\end{equation}}
\newcommand{\bea}{\begin{eqnarray}}
\newcommand{\eea}{\end{eqnarray}}
\newcommand{\ba}{\begin{eqnarray*}}
\newcommand{\ea}{\end{eqnarray*}}
\newcommand{\dagga}{{\phantom{\dagger}}}
\newcommand{\bR}{\mathbf{R}}
\newcommand{\bQ}{\mathbf{Q}}
\newcommand{\bq}{\mathbf{q}}
\newcommand{\bqp}{\mathbf{q'}}
\newcommand{\bk}{\mathbf{k}}
\newcommand{\bh}{\mathbf{h}}
\newcommand{\bkp}{\mathbf{k'}}
\newcommand{\bp}{\mathbf{p}}
\newcommand{\bL}{\mathbf{L}}
\newcommand{\bRp}{\mathbf{R'}}
\newcommand{\bx}{\mathbf{x}}
\newcommand{\by}{\mathbf{y}}
\newcommand{\bz}{\mathbf{z}}
\newcommand{\br}{\mathbf{r}}
\newcommand{\Ima}{{\Im m}}
\newcommand{\Rea}{{\Re e}}
\newcommand{\Pj}[2]{|#1\rangle\langle #2|}
\newcommand{\ket}[1]{\vert#1\rangle}
\newcommand{\bra}[1]{\langle#1\vert}
\newcommand{\setof}[1]{\left\{#1\right\}}
\newcommand{\fract}[2]{\frac{\displaystyle #1}{\displaystyle #2}}
\newcommand{\Av}[2]{\langle #1|\,#2\,|#1\rangle}
\newcommand{\av}[1]{\langle #1 \rangle}
\newcommand{\Mel}[3]{\langle #1|#2\,|#3\rangle}
\newcommand{\Avs}[1]{\langle \,#1\,\rangle_0}
\newcommand{\eqn}[1]{(\ref{#1})}
\newcommand{\Tr}{\mathrm{Tr}}

\newcommand{\Vb}{\bar{\mathcal{V}}}
\newcommand{\Vd}{\Delta\mathcal{V}}
\def\P{P_{02}}
\newcommand{\Pb}{\bar{P}_{02}}
\newcommand{\Pd}{\Delta P_{02}}
\def\t{\theta_{02}}
\newcommand{\tb}{\bar{\theta}_{02}}
\newcommand{\td}{\Delta \theta_{02}}
\newcommand{\Rb}{\bar{R}}
\newcommand{\Rd}{\Delta R}
\newcommand{\ocrev}[1]{{\color{cyan}{#1}}}
\newcommand{\occom}[1]{{\color{red}{#1}}}

%\title{Combining quantum-embedding theories with machine learning}

%%%\title{Bypassing the computational bottleneck of quantum-embedding methods with machine learning}

\title{Bypassing the computational bottleneck of quantum-embedding theories for \\ strong electron correlations with machine learning}

\author{John Rogers}
\affiliation{Department of Physics and Astronomy, Texas A\&M University, College Station, Texas 77845, USA}
\author{Tsung-Han Lee}
\affiliation{Physics and Astronomy Department, Rutgers University, Piscataway, New Jersey 08854, USA}
\author{Sahar Pakdel}
\affiliation{Department of Physics and Astronomy, Aarhus University, 8000,
Aarhus C, Denmark}
\author{Wenhu Xu}
\affiliation{Condensed Matter Physics and Materials Science Department, Brookhaven National Laboratory, Upton, NY 11973}
\author{Vladimir Dobrosavljevi\'c}
\affiliation{Department of Physics and National High Magnetic Field Laboratory, Florida State University, Tallahassee, Florida 32306, USA}
\author{Yong-Xin Yao}
%\altaffiliation{Corresponding author: ykent@iastate.edu}
\affiliation{Ames Laboratory-U.S. DOE and Department of Physics and Astronomy, Iowa State University, Ames, Iowa 50011, USA}
\author{Ove Christiansen}
\altaffiliation{Corresponding author: ove@chem.au.dk}
\affiliation{Department of Chemistry, Aarhus University, 8000, Aarhus C, Denmark}
\author{Nicola Lanat\`a}
\altaffiliation{Corresponding author: lanata@phys.au.dk}
\affiliation{Department of Physics and Astronomy, Aarhus University, 8000, Aarhus C, Denmark}

\date{\today}

\begin{abstract}

A cardinal obstacle to performing quantum-mechanical simulations of strongly-correlated matter is that, with the theoretical tools presently available, sufficiently-accurate computations are often too expensive to be ever feasible.
Here we design a computational framework combining quantum-embedding (QE) methods with machine learning.
This allows us to bypass altogether the most computationally-expensive  components of QE algorithms, making their overall cost comparable to bare Density Functional Theory (DFT).
We perform benchmark calculations of a series of actinide systems, where our method describes accurately the correlation effects, reducing by orders of magnitude the computational cost.
We argue that, by producing a larger-scale set of training data, it will be possible to apply our method to systems with arbitrary stoichiometries and crystal structures,
paving the way to virtually infinite applications in condensed matter physics, chemistry and materials science.

\end{abstract}

\maketitle

\section{Introduction}

The atomic energy scales emerging in “strongly correlated” systems~\cite{Mott-book,Kotliar-Science,Adler_2018} 
can induce a broad spectrum of spectacular effects, ranging from arresting the electronic motion~\cite{Mott-book} to causing high-temperature superconductivity~\cite{highTc-1}, 
%%~\cite{highTc-1,highTc-2,highTc-3,highTc-4,highTc-5,highTc-6,highTc-7,highTc-8},
unlocking access to new topological phases
%~\cite{corr-topology-2,corr-topology-3,corr-topology-4,corr-topology-5,corr-topology-6,corr-topology-7}
%~\cite{corr-topology-1,corr-topology-2,corr-topology-3,corr-topology-4,corr-topology-5,corr-topology-6,corr-topology-7}
and influencing dramatically the potential-energy surfaces (PES) of molecules and solids~\cite{corr-structure-1,corr-structure-3,corr-structure-4,corr-structure-5,dmft_forces,npj-lanata}. 
%~\cite{corr-structure-1,corr-structure-2,corr-structure-3,corr-structure-4,corr-structure-5,corr-structure-6,dmft_forces,Our-PRX,npj-lanata}. 
Therefore, the need and the potential effects for science and society of extending to strongly correlated systems  the computational materials-by-design paradigm can hardly be overstated~\cite{Kotliar-Science}.
The substantial progress achieved in the past decade in calculating
the electronic structure of strongly correlated materials
largely owes to the idea of combining mean-field (MF) theories, such as
approximations to DFT~\cite{HohenbergandKohn,KohnandSham,LDA,GGA-simple,DFT-RevModPhys.87.897,DFT-doi:10.1080/00268976.2017.1333644,DFT-doi:10.1080/00268976.2017.1333644,DFT-doi:10.1063/1.4704546}
%~\cite{HohenbergandKohn,KohnandSham,LDA,GGA-simple,DFT-RevModPhys.87.897,DFT-annurev-physchem-052516-044835,DFT-doi:10.1080/00268976.2017.1333644,DFT-doi:10.1080/00268976.2017.1333644,DFT-doi:10.1063/1.4704546}, 
with QE~\cite{quantum-embedding-review,Kotliar-Science, quantum-embedding-another} 
theoretical frameworks.
Well-known examples 
are Dynamical Mean Field Theory (DMFT)~\cite{DMFT,dmft_book,LDA+U+DMFT,Held-review-DMFT,Anisimov_DMFT,CDMFT-Jarrell,CDMFT-Potthoff,CDMFT-Lichtenstein} and Density Matrix Embedding Theory (DMET)~\cite{DMET,Bulik-DMET}.
As shown in Ref.~\cite{Our-PRX},
also the multi-orbital Gutzwiller approximation
(GA)~\cite{Gutzwiller3,Gebhard,Fang,Our-PRX,Ghost-GA}, which is a variational framework (equivalent to the Rotationally Invariant Slave Boson (RISB)~\cite{Fresard1992,Georges,Lanata2016}
at the MF level~\cite{equivalence_GA-SB,lanata-barone-fabrizio}),
can be formulated as a QE scheme
featuring recursive ground-state calculations of impurity models with a finite bath called "Embedding Hamiltonians" (EH). 
Therefore, even if the principles underlying E=DMFT,GA,RISB,DMET are very different, the concept of QE
allows us to formalize and implement these
techniques from a unified perspective~\cite{Our-PRX,dmet-risb-1,dmet-risb-2}.

The fundamental idea underlying  all QE theoretical frameworks consists in
replacing the original (typically unfeasible) problem of directly simulating these systems with the
more manageable task of solving equations for a series of EHs,
composed by
fractions of the material (impurities) and 
effective-medium
degrees of freedom 
(self-consistently determined
for describing the interaction of the impurities with their
environment).
The current state-of-the-art approach to tackle QE simulations
is based on solving the EH equations recursively 
%consists in solving recursively the EH equations 
utilizing many-body techniques~\cite{many-qmc,Wilson,DMRG-original-White-PRL,Vaestrate_NRG-DMRG-MPS}.
%~\cite{many-qmc,afqmc-1,afqmc-2,ctqmc,ctqmc_hybr-exp_Rubtsov,ctqmc-RevModPhys.83.349,ctqmc-PhysRevB.90.075149,Hewson,Wilson,DMRG-original-White-PRL,DMRG-original-White-PRB,PRB-DMRGvsNRG,Vaestrate_NRG-DMRG-MPS}.
%
On the other hand, due to the
quantum-mechanical nature of the EH, its solution ultimately 
has a computational
cost that grows exponentially with the number of impurity degrees of freedom.
Because of this reason, the practical application of these tools to complex materials is often too computationally demanding to be ever feasible.

Here we show that this problem can be efficiently tackled
 from a completely different perspective:
capitalizing on the fact that the form of the EH is \emph{universal}
(i.e., it does not depend on the specific stoichiometry and
crystal structure of the material considered), we 
bypass altogether the computationally-expensive 
recursive solution of the EH
by "training a machine" to solve this problem \emph{once and for all}.
To accomplish this goal, we develop a computational framework
combining machine-learning (ML) techniques, such as “Kernel Ridge Regression” (KRR), with a mathematical method named ``$n$-mode representation''~\cite{n-mode-1,n-mode-2} ---previously used 
for effectively reducing the dimensionality of large-scale regression problems (e.g., for reducing the number of points required for constructing high-dimensional PESs in quantum chemistry~\cite{n-mode-3,n-mode-4}).

Note the fundamentally-different nature of our method
---which employs ML \emph{inside} the solution of the full quantum problem,--- with respect to the many current uses of ML for learning pre-existing solutions~\cite{ML-Schmidt2019} (e.g., for applications to different materials and structures).

We illustrate the power of our method 
by performing benchmark 
calculations of a series of actinides.
Utilizing our method, we were able to calculate
---at a computational cost comparable to bare DFT--- 
the discontinuous behavior of the
equilibrium volumes of the actindes as a function
of their atomic number Z
(actinide transition)~\cite{Pu-nature-albers-2001},
which is a phenomenon originated by a complex interplay between
structural degrees of freedom, relativistic effects,
atom- and orbital- selective electron correlations~\cite{Our-PRX,Amadon-actinides, alphaPu-Gabi-nature-2013,Pu-Gabi-nature-2001,sh-alpha-delta-Pu}.
%~\cite{Our-PRX,Amadon-actinides, alphaPu-Gabi-nature-2013,Pu-Gabi-nature-2001,sh-alpha-delta-Pu,deltaPu-Sensitivity-nf-spectral,deltaPu-ImportanceFullCoulomb-1,deltaPu-ImportanceFullCoulomb-2}

\begin{figure} %[htp]
    \includegraphics[width=8.6cm]{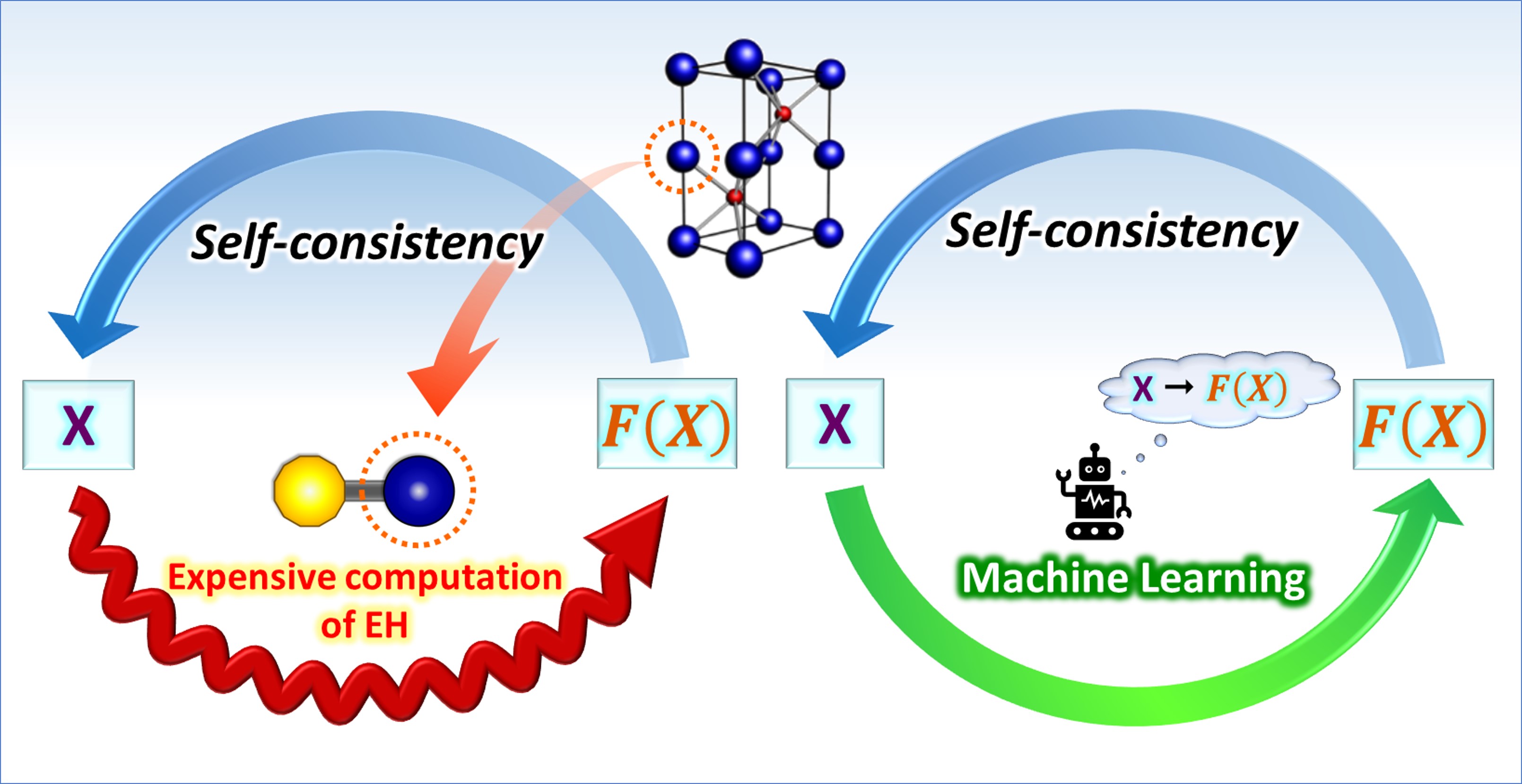}
    \caption{Algorithmic structure of QE implementations. Standard approach (left): each iteration requires to solve numerically the EH and calculate the observables $F(X)$ for different descriptors $X$. Proposed approach (right): a ML algorithm, previously trained to learn the universal function $F(X)$, allows us to bypass the computation of the EH.}
    \label{Figure1}
\end{figure}

\section{QE algorithmic structure}

The problem of applying QE methodologies (such as the DFT+E methods) to realistic solids and molecules ultimately reduces to solving recursively
multi-orbital Hamiltonians represented as follows:
\be
\h= 
\sum_{\bk} 
\sum_{ij=1}^\eta 
\sum_{\alpha=1}^{M_i} 
\sum_{\beta=1}^{M_j} 
\epsilon^{\alpha\beta}_{\bk,ij}
\,\cc_{\bk i\alpha}\ca_{\bk j\beta}
+\sum_{\bR i}\h^{\text{loc}}_{\bR i}
\,;\label{hubbard}
\ee
where $\bk$ is the momentum conjugate to the unit-cell label $\bR$,
the electronic shells of the atoms within the unit cell are labeled by $i,j=1,..,\eta$,
and the corresponding spin-orbitals are labeled by $\alpha=1,..,M_i$,
$\beta=1,..,M_j$.
For later convenience, with no loss of generality,
we assume that the first term is non-local (i.e, that
$\sum_k\epsilon_{\bk,ii}=0\;\forall\, i$)
and that $\h^{\text{loc}}$ includes
both the one-body and the two-body local parts of $\h$:
\be
\h^{\text{loc}}_{\bR i}=\sum_{\alpha\beta=1}^{M_i}
[E_{i}]_{\alpha\beta}\,\cc_{\bR i\alpha}\ca_{\bR i\beta} +
\h^{\text{int}}_{\bR i}[U_i,J_i]\,,
\label{impurity}
\ee
where $E_{i}$ describe the on-site energies
(such as the crystal-field energies and the spin-orbit coupling (SOC))
and $\h^{\text{int}}_{\bR i}$ depends on the
Slater-Condon parameters~\cite{LDA+U}, i.e., 
the Hubbard interaction
strength $U_i$ and the Hund's coupling constant $J_i$.

The basic algorithmic structure of all QE methods to solve
the Hamiltonian
[Eq.~\eqref{hubbard}]
is schematically illustrated in the left side of Fig.~\ref{Figure1}.
A series of EH, represented as:
\be
\h_{\bR i}^{\text{emb}}[U_i,J_i, E_i, x_i]=
\h^{\text{loc}}_{\bR i}[U_i,J_i,E_i]
+\hat{\mathcal{B}}_{\bR i}(x_i)\,,
\label{hemb-abstract}
\ee
are constructed 
%in order to describe 
for describing the coupling of the impurity with its environment in
a MF fashion. 
Here $\hat{\mathcal{B}}_{\bR i}(x_i)$ represents an effective medium coupled with the subsystem (impurity) [Eq.~\eqref{impurity}], which is encoded in a series of parameters $x_i$.
Determining the self-consistent parameters $x_i$
requires to calculate multiple times a series of 
quantities (varying for different QE methodologies) for the Hamiltonian in Eq.~\eqref{hemb-abstract}, that we schematically represent as $F_i$.

For concreteness, here we focus on the GA.
As shown in Refs.~\cite{Our-PRX,Lanata2016} (see also the supplemental material), this method can be regarded as a QE framework where:
\be
\hat{\mathcal{B}}_{\bR i}=
\sum_{a \alpha=1}^{M_i} \left(
\left[\D_{i}\right]_{a\alpha}
{c}^\dagger_{i \alpha}{f}^\dagga_{i a}+\text{H.c.}\right)
+\sum_{a b=1}^{M_i} \left[\lambda^c_{i}\right]_{ab}
{f}^\dagga_{i b}{f}^\dagger_{i a}\,,
\label{h-emb-i}
\ee
$\D_i$, $\lambda_i^c$ are complex $M_i\times M_i$ matrices, the latin labels $a,b$ correspond to the bath degrees of freedom $f$ and the output
function $F_i$ is the single-particle density matrix:
\begin{align}
[F_{i}]_{AB}&=\langle\Phi_i| [{\psi}^\dagger_{i}]_{A} [{\psi}^\dagga_{i}]_{B} |\Phi_i\rangle
\label{rho}
\\
{\psi}_{i}&=\big(
{c}_{i 1},..,{c}_{i M_i},
{f}_{i 1},..,{f}_{i M_i}
\big)\,,
\end{align}
where the labels $A,B=1,..,2M_i$ run over both the impurity and the
bath degrees of freedom.
Therefore,
consistently with the general algorithmic structure schematically represented in the left side of
Fig.~\ref{Figure1}, solving the GA equations requires to evaluate recursively Eq.~\eqref{rho}
as a function of the EH descriptors:
\begin{align}
X_i&=(U_i,J_i, E_i, x_i)
\label{XX}
\\
x_i&=(\D_i,\lambda_i^c)\,.
\label{xx}
\end{align}

%%A brief overview of the key equations underlying the QE formulation of the GA is provided in the supplemental material.

To simplify the notation, from now on we will omit the electronic-shell label $i$.
%in the function $F_i(X_i)$ of Eq.~\eqref{rho}.

\subsection*{Computational complexity of the EH problem}

In GA ab-initio calculations %of real materials,
it is typically necessary to deal with impurities 
consisting of $M=10$ degrees of freedom (for d-electron shells) or $M=14$ degrees of freedom
(for f-electron shells).
Since the bath of the EH has the same number of degrees of freedom of the impurity, see Eq.~\eqref{h-emb-i},
the dimension of the EH space is $D=2^{2M}$.

Note that the dimension of the EH system scales as $D=2^{2M}$ also in DMET.
In fact, the differences between these 2 methods stem exclusively
from their different self-consistency conditions~\cite{Our-PRX,dmet-risb-1,dmet-risb-2}.
Within the ghost GA framework (g-GA), which is a more accurate extension of the GA~\cite{Ghost-GA},
the number of effective-medium degrees of freedom is still finite, but larger than bare GA. Therefore, the dimension $D$ of the EH system is exponentially higher.
Finally, in DMFT~\cite{DMFT} the number of effective-medium degrees of freedom (and, therefore, the EH dimension $D$) is infinite.

In all of the theoretical tools mentioned above, the computational bottleneck is solving recursively the EH equations. In fact, this is the only reason why the cost of QE methods generally exceeds by orders of magnitude the cost of mean-field approaches, such as classic approximations to DFT.
The computational framework described in the next section will allow us to bypass altogether this problem.

\section{Combining ML with the ${n}$-mode representation}\label{3}

Rather than trying to develop more efficient computational tools for solving the EH equations,
in this work we will capitalize on the universality
of the function [Eq.~\eqref{rho}],
utilizing KRR and the $n$-mode expansion for learning it
%to learn it %Eq.~\eqref{rho} 
once and for all, see the right side of Fig.~\ref{Figure1}.

The strategy of utilizing ML for bypassing expensive calculations of universal maps, 
often referred to as ``surrogate modeling,''
is widely used in physics, chemistry and materials
science~\cite{Kanungo2019,surrogate-1,surrogate-2,surrogate-DFT-1,surrogate-DFT-2,surrogate-Millis,ML-materials-1}.
The main obstacle to applying classic
ML algorithms (such as KRR) for learning multi-variable functions is that it requires a number of training data points that scales as:
\be
N\sim m^d\,,
\label{exponential-scaling}
\ee
where $d$ is the number of input variables and $m$ is the number of mesh subdivisions for each dimension.
This problem is often referred to as the ``exponential curse''.

Without use of symmetry,
the number of input variables in $F(X)$ in Eq.~\eqref{rho} is
$d=1+4M^{2}$, which is $401$ for d-electron shells and $785$ for f-electron shells.
Therefore, direct applications of ML methods
would be extremely costly
%are generally insufficient 
for learning this function.

To overcome this problem, here we combine
KRR with the ``$n$-mode representation,'' which is 
a technique previously explored in different contexts (and under different names), e.g., for reducing the
number of points required to construct
high-dimensional PESs~\cite{jung_vibrational_1996,n-mode-1,n-mode-3,n-mode-4} and for facilitating the solution of the Schr{\"o}dinger equation in quantum chemical methods~\cite{zimmerman_strong_2017,stoll_toward_2019}.
The basic idea underlying the $n$-mode representation is to construct approximations to the high-dimensional function $F(X)$
in terms of the so-called
``cut-functions,'' such as:
\begin{align}
F^0 &=  F(0,0,\dots,0,0,0,\dots,0,0,0,\dots,0)
\label{cut}
\\ 
F^{1}_i(X_i) &=  F(0,0,\dots,0,X_i,0,\dots,0,0,0,\dots,0)
\nonumber\\ 
F^{2}_{ij}(X_i,X_j) &=  F(0,0,\dots,0,X_i,0,\dots,0,X_j,0\dots,0)
\nonumber
\,,
\end{align}
which are the restrictions of $F(X)$ to hyperplanes where subsets of the components of $X$ are set to $0$.
At a given order $n$ of the expansion, $F(X)$ is approximated utilizing only cut functions of up to $n$ variables.
The exact function is recovered when $n$
equals the total number of variables $d$, and the series converges very rapidly as a function of $n$ in many relevant cases~\cite{jung_vibrational_1996,n-mode-1,n-mode-3,n-mode-4,zimmerman_strong_2017,stoll_toward_2019}.
%
%%%More details about the $n$-mode expansion are provided in the supplemental material.

\begin{figure} %[htp]
    \includegraphics[width=8.6cm]{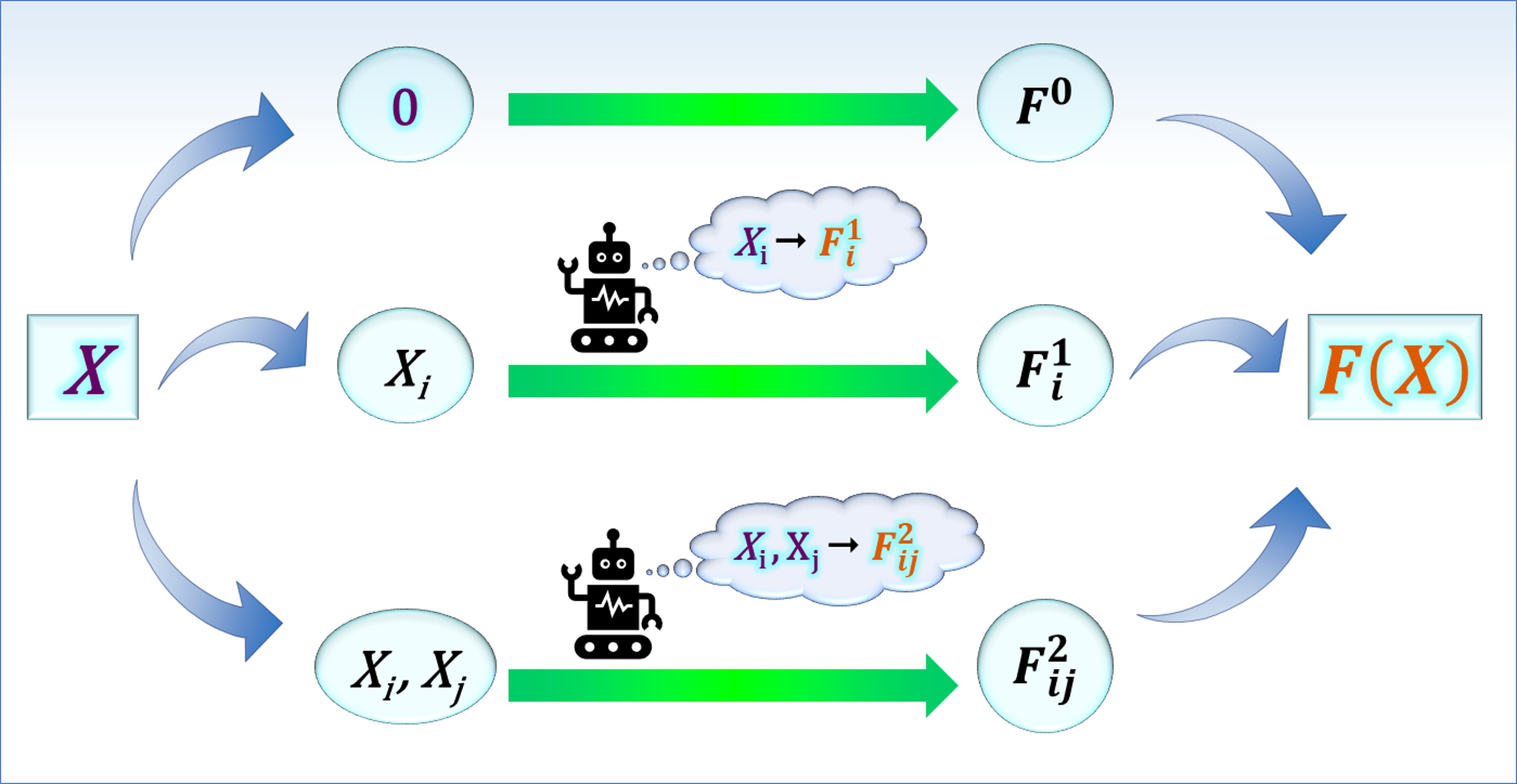}
    \caption{Representation of the proposed nKRR approach. The universal EH functions $F(X)$ is approximated 
    with the $n$-mode representation
    up to the desired order $n$ (e.g., $n=2$ in the picture). 
    The KRR method is used to fit the corresponding
    lower-dimensional cut-functions, which are recombined into
    an approximation to $F(X)$.}
    \label{Figure2}
\end{figure}

Within our context of application, the main consequence of the $n$-mode expansion is that, since the effective dimensionality is limited to that of the needed order $n$, the input-output mapping can be determined 
with a number of data points that scales only as:
\be
N^{(n)}\sim {d}^n\,,
\label{polynomial-scaling}
\ee
i.e., it scales polynomially as a function of $d$, rather than exponentially (Eq.~\eqref{exponential-scaling}).
%is exponentially smaller than Eq.~\eqref{exponential-scaling} for $n<d$.
%
Specifically, the hereby
proposed nKRR methodology consists of the following steps:
\begin{itemize}
\item Breaking down the universal functions
$F(X)$ in lower-dimensional cut-functions,
using the $n$-mode expansion.
\item Learning the corresponding lower-dimensional cut functions, up to the desired order, using KRR.
\item Combining the cut-functions into the desired
$n$-mode approximation.
\end{itemize}
A schematic representation of this approach is shown in
Fig.~\ref{Figure2}.
For completeness, a short introduction to the KRR method and the $n$-mode representation is provided in the supplemental material.

We point out that, as opposed to other dimensionality-reduction techniques (where the number of input variables is decreased), the nKRR method allows us to take into account from the outset
\emph{all} descriptors of the EH system ---in a manner such that the effective dimensionality is substantially reduced.
In the next section 
we will also capitalize on general physical arguments inherent in the specific
structure of Eq.~\eqref{h-emb-i}. This will allow us to derive a convenient parametrization of the EH in DFT+GA calculations, 
speeding up dramatically the convergence of the $n$-mode representation.

We want to point out that, besides the GA, the nKRR methodology
described above could as well
be implemented in combination with DMET or more accurate QE methods, such as the g-GA~\cite{Ghost-GA} and DMFT.

%%\section{Implementation of the $\text{n}$KRR method for actinide systems}\label{4}

\section{Application to actinide systems}\label{4}

Here we describe in detail our implementation of the nKRR
method for actinide systems.
For simplicity, we will focus
on the case of a generic EH consisting of f-electron shells in an isotropic
medium ---which is typically a good approximation for
actinide systems, where the dominant role of 
the SOC allows us to 
average over the crystal-field splittings.
Under these assumptions, using group-theoretical considerations, it can be shown~\cite{Lanata2016,Gmethod}
that the $14\times 14$ matrices
$E,\D,\lambda^c$
are diagonal and fully determined by their respective $j=5/2$ and $j=7/2$
components $E_j,\D_j,\lambda^c_j$, where $j$ is the label of the total angular
momentum for an f-electron shell.
Furthermore, as discussed in the supplemental material,
the conservation of the total number of electrons implies that the trace of the single-particle density matrix of the EH
is $M$.
Therefore, 
the only independent descriptors of the EH
are the interaction parameters $U,J$ and the following variables:
\begin{align}
   X_1&=\frac{1}{4}\left(E_{5/2}+E_{7/2}+\lambda^c_{5/2}+\lambda^c_{7/2}\right)
   \nonumber\\
   X_2&=\frac{1}{2}\left(E_{5/2}-E_{7/2}\right)
   \nonumber\\
   X_3&=\frac{1}{2}\left(\lambda^c_{5/2}-\lambda^c_{7/2}\right)
   \nonumber\\
   X_4&=\D_{5/2}
   \nonumber\\
   X_5&=\D_{7/2}
   \,,
   \label{xi}
\end{align}
In fact, with no loss of generality, we can set $\sum_{j\in\{5/7,7/2\}}(E_j-\lambda^c_j)=0$, as changing this variable corresponds to applying a chemical-potential shift in Eq.~\eqref{h-emb-i}
(which would be redundant, as the number of electrons $M$ in the EH is fixed).

Furthermore, the behavior of $F(X)$ (Eq.~\eqref{rho}) is fully
determined by the following functions:
\begin{align}
   F_1&=\frac{1}{4}\sum_{j=\frac{5}{2},\frac{7}{2}}
   \sum_{j_z=-j}^{j}
   \left(\langle\hat{c}^\dagger_{jj_z}\hat{c}^\dagga_{jj_z}\rangle
   -\langle\hat{f}^\dagger_{jj_z}\hat{f}^\dagga_{jj_z}\rangle
   \right)
   \nonumber\\
   F_2&=\frac{1}{2}\left(
   \sum_{j_z=-\frac{5}{2}}^{\frac{5}{2}}
   \langle\hat{c}^\dagger_{\frac{5}{2}j_z}\hat{c}^\dagga_{\frac{5}{2}j_z}\rangle
   -\sum_{j_z=-\frac{7}{2}}^{\frac{7}{2}}
   \langle\hat{c}^\dagger_{\frac{7}{2}j_z}\hat{c}^\dagga_{\frac{7}{2}j_z}\rangle
   \right)
   \nonumber\\   
   F_3&=\frac{1}{2}\left(
   \sum_{j_z=-\frac{5}{2}}^{\frac{5}{2}}
   \langle\hat{f}^\dagger_{\frac{5}{2}j_z}\hat{f}^\dagga_{\frac{5}{2}j_z}\rangle
   -\sum_{j_z=-\frac{7}{2}}^{\frac{7}{2}}
   \langle\hat{f}^\dagger_{\frac{7}{2}j_z}\hat{f}^\dagga_{\frac{7}{2}j_z}\rangle
   \right)
   \nonumber\\  
   F_4&=\frac{1}{2}\left(
   \sum_{j_z=-\frac{5}{2}}^{\frac{5}{2}}
   \langle\hat{c}^\dagger_{\frac{5}{2}j_z}\hat{f}^\dagga_{\frac{5}{2}j_z}\rangle
   +
   \sum_{j_z=-\frac{7}{2}}^{\frac{7}{2}}
   \langle\hat{c}^\dagger_{\frac{7}{2}j_z}\hat{f}^\dagga_{\frac{7}{2}j_z}\rangle
   \right)
   \nonumber\\  
   F_5&=\frac{1}{2}\left(
   \sum_{j_z=-\frac{5}{2}}^{\frac{5}{2}}
   \langle\hat{c}^\dagger_{\frac{5}{2}j_z}\hat{f}^\dagga_{\frac{5}{2}j_z}\rangle
   -\sum_{j_z=-\frac{7}{2}}^{\frac{7}{2}}
   \langle\hat{c}^\dagger_{\frac{7}{2}j_z}\hat{f}^\dagga_{\frac{7}{2}j_z}\rangle
   \right)
   \,,
   \label{fi}
\end{align}
where $j_z$ is the quantum label of the
third component of the total angular momentum for each $j$.

Consistently with Ref.~\cite{Our-PRX}, here we set the screened Hubbard interaction $U=4.5$~eV and the Hund's coupling
constant $J=0.36$~eV. Therefore, the only free embedding descriptors are $X_1,..,X_5$.

\subsection*{Parametrization of $F(X)$ and training data set} %\label{setup}

In this section we describe in detail our procedure for setting up the nKRR method, that we are going to utilize for performing DFT+GA calculations of systems involving Pa, U, Np, Pu and Am.

%%%

As pointed out in the supplemental material, the speed of convergence of the $n$-mode representation
can be improved
by a suitable change of variables, as (by construction) the accuracy of the approximation tends to be higher in the proximity of the domain of the cut functions utilized at the chosen order of truncation.
Furthermore, it is pivotal to ensure that the grid of training data points utilized for learning the cut functions [Eq.~\eqref{cut}] is sufficiently large, as ML methods can be predictive only within the training-data range.

%Within our context of application, 
It is particularly convenient for our applications to exploit the
possibility of expressing the function  of
Eq.~\eqref{fi} 
in terms of shifted variables:
\be
Y=X-\bar{X}\,.
\label{yi}
\ee
For setting up the components of $\bar{X}$ and the training data set we utilized the following procedure, which is based on physical considerations inherent in the properties of the atomic impurities of interest.

\begin{figure} 
    \includegraphics[width=8.8cm]{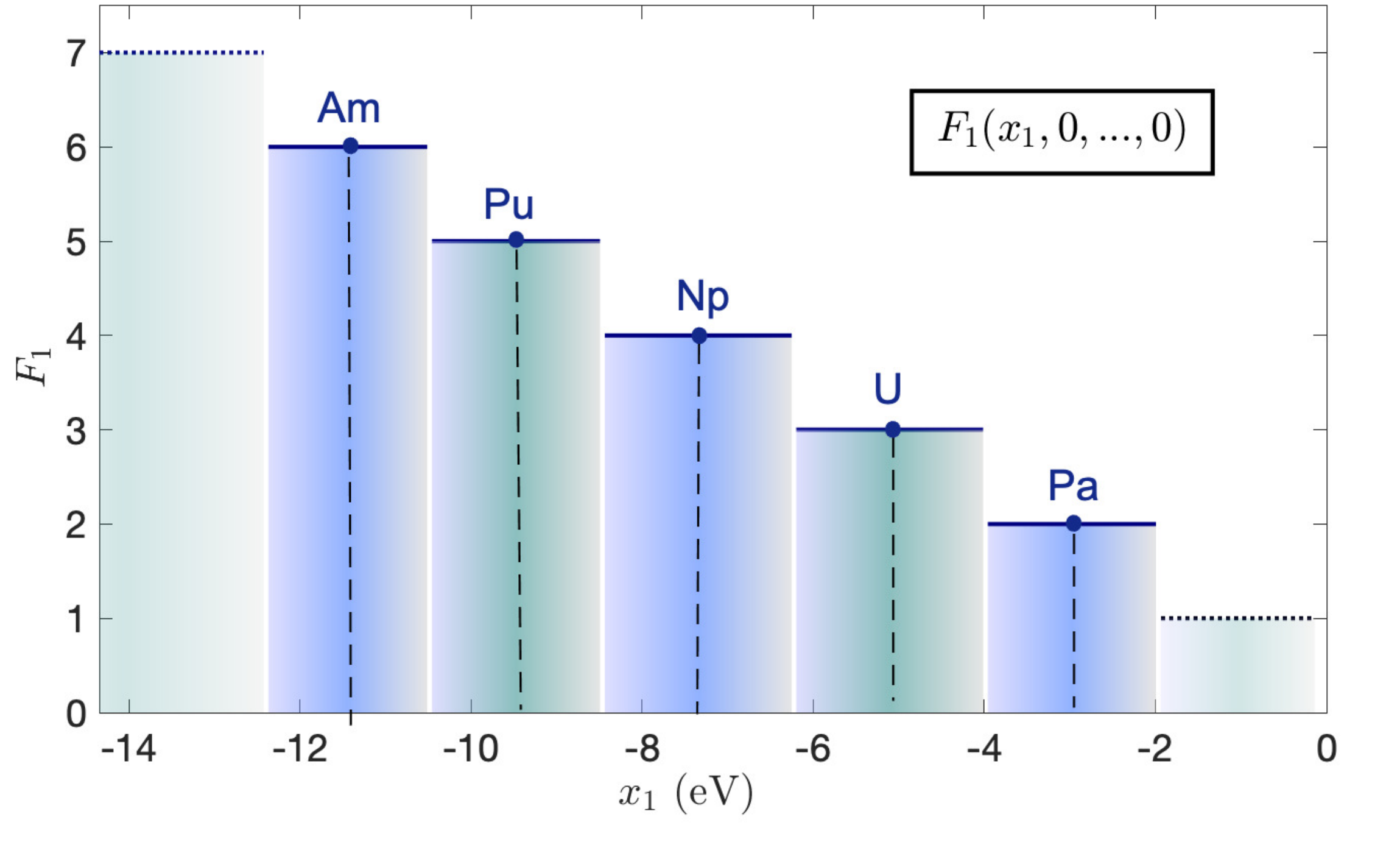}
    \caption{Behavior of $F_1(X_1,0,..,0)$, representing the  occupation of the impurity for an EH disentangled from the bath site. The mid values of $X_1$ of the plateaus are used to set the values of
    $\bar{x}_1$ for all actinides considered.}
    \label{Figure3}
\end{figure}

To determine $\bar{X}_1$,
we have pre-calculated the behavior of $F_1$ as a function of  $X_1$, at fixed
$X_j=0$ $\forall\,j=2,..,5$.
Note that, in this limit, the impurity
$\h^{\text{loc}}$ is isolated from the bath. 
Therefore, the values of $F_1$ are quantized and correspond to the nominal (integer) f-electron occupations along the actinide series.
For each actinide, we have set $\bar{X}_1$
as the middle point of the interval of $X_1$ values such that $F_1(X_1,0,..,0)$
equals the corresponding nominal occupation, see Fig.~\ref{Figure3}. % and Table~\ref{table:z}.
We note that ${X}_2$ describes the impurity SOC, which is essentially an atomic property, i.e., it is typically almost independent of the environment in DFT and DFT+GA calculations.
Therefore, for each actinide
we have set $\bar{X}_2$ based on the to the nominal atomic values, % reported in Table~\ref{table:z}, 
which we pre-calculated using LDA.
Since $X_3$, $X_4$, $X_5$ describe the EH bath and its  coupling with the impurity, their range
is generally system dependent. Therefore,
the choice of $\bar{X}_3$, $\bar{X}_4$, $\bar{X}_5$ is essentially arbitrary. 
In our calculations we have set them based on a single DFT+GA calculation of $\delta$-Pu at its experimental equilibrium volume.

The range of the training data set was estimated by performing
LDA+GA calculations of $\delta$-Pu at $\pm$35\%
of its experimental equilibrium volume. 
This choice proved to be sufficient for performing all calculations performed in this work.
Note that our implementation interactively queries the user if EH parameters beyond the training range are explored in a calculation.
Whenever this happens,
new training data can be generated and stored in a database.
This type of iterative supervised learning ---which is often called "active learning procedure,"--- allows one to assess the validity of the simulations and to extend systematically the range of applicability of the nKRR algorithm.

The numerical values of the components of $\bar{X}$ and the data mesh of the components of $Y$ ---obtained with the procedure outlined above--- are reported in the supplemental material.

\begin{figure*}[t]
    \includegraphics[width=18cm]{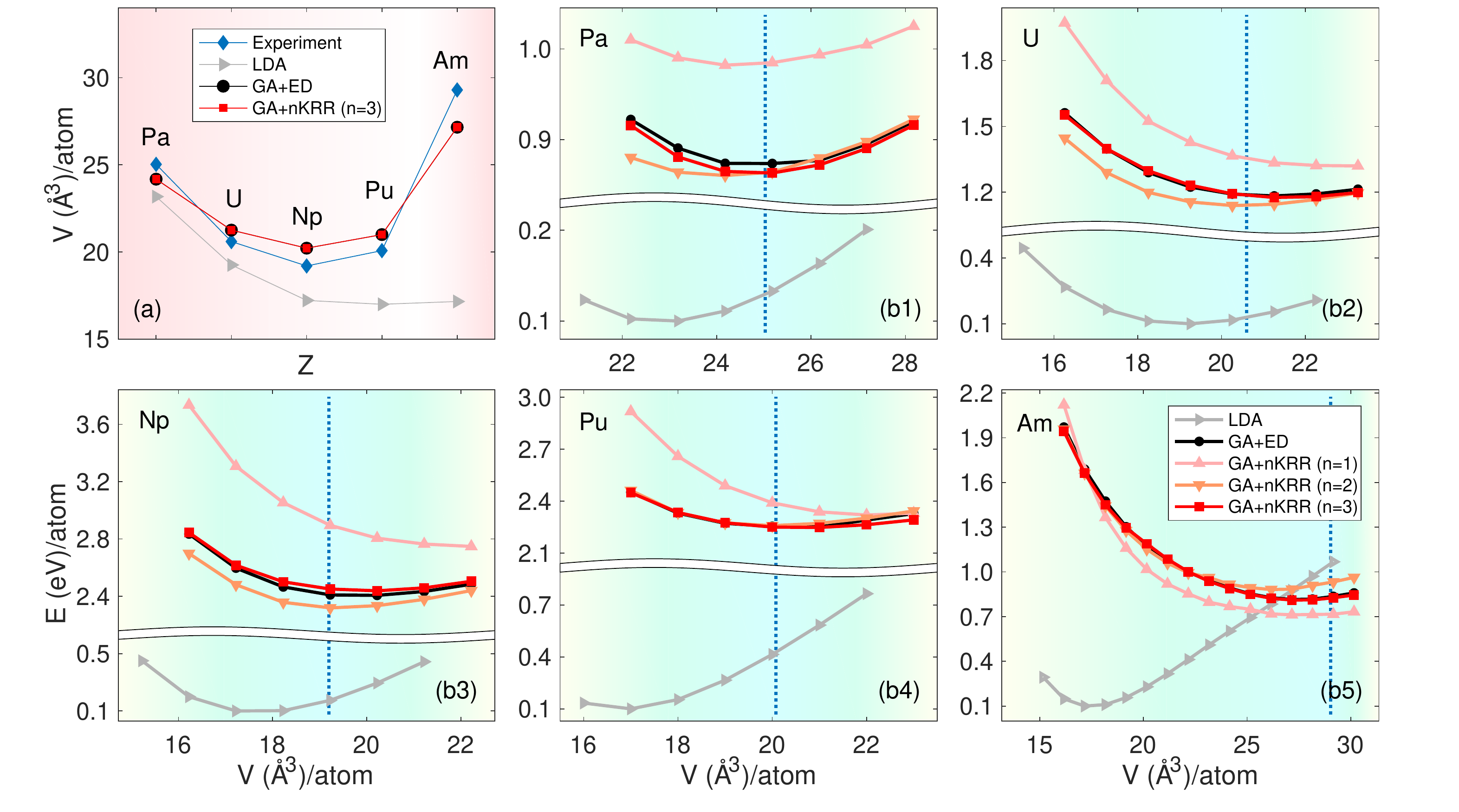}
    \caption{
    \emph{Panel (a):} GA+nKRR ($n=3$) and GA+ED equilibrium volumes of the the low-temperature allotropes of Pa, U, Np, Pu, Am; in comparison with bare LDA and the experimental values.
    \emph{Panels (b1-b5):} Corresponding LDA+GA+nKRR ($n=1,2,3$), GA+ED and LDA energy-volume curves.
    The vertical blue dashed lines indicate the experimental equilibrium volumes.
    Within the discrete mesh of volumes considered, the GA+nKRR ($n=3$) and GA+ED minima correspond to the same equilibrium points (panel (a)).
    }
    \label{figure4}
\end{figure*}

\section{Benchmark calculations}

To assess the power of our method, we performed LDA+GA benchmark calculations 
of different actinide solids, 
utilizing the nKRR method described above to solve the 
EH Hamiltonian.
%
%For clarity, below w
We will refer to this framework as the GA+nKRR, while we will call GA+ED the standard DFT+GA approach resulting from using ED as an EH solver.

A particularly interesting property of the actinide series is
the anomalous dependence of their equilibrium volumes as a function of the atomic number.
In fact, while the equilibrium volume of the lighter actinides
varies continuously as a function $Z$ (from Pa to Pu),
it displays a pronounced discontinuity between Pu and Am.
This volume anomaly is often called
\emph{actinides transition}~\cite{Pu-nature-albers-2001,Actinide_metals-review},
and it is originated by a complex interplay between
structural degrees of freedom with SOC,
atom- and orbital- selective electron
correlations~\cite{Our-PRX,Amadon-actinides, alphaPu-Gabi-nature-2013,Pu-Gabi-nature-2001,sh-alpha-delta-Pu}.
%~\cite{Our-PRX,Amadon-actinides, alphaPu-Gabi-nature-2013,Pu-Gabi-nature-2001,sh-alpha-delta-Pu,deltaPu-Sensitivity-nf-spectral,deltaPu-ImportanceFullCoulomb-1,deltaPu-ImportanceFullCoulomb-2}.
%
Therefore, capturing this behavior constitutes a very strict benchmark of our method.

In panel Fig.~\ref{figure4}(a) we show the equilibrium volumes of the low-temperature allotropes of Pa, U, Np, Pu and Am. 
The GA+nKRR ($n=3$) and GA+ED calculations 
(performed at $U=4.5$~eV and $J=0.36$~eV, as in Ref.~\cite{Our-PRX},)
are shown in comparison with LDA and the experiments~\cite{Amadon-actinides,Zachariasen-Pa}.
In Fig.~\ref{figure4}(b1-b5) we also show the corresponding energy-volume curves ---whose minima are the points shown in Fig.~\ref{figure4}(a).
Remarkably, for all systems considered, the GA+nKRR method is substantially more accurate than DFT already for $n=2$, while it becomes essentially as accurate as GA+ED for $n=3$.

We point out that each ED solution of the EH Hamiltonian
takes about 10 minutes and requires 10GB of RAM on average, while it takes only about 0.1 seconds and 50MB of RAM within the nKRR framework.
Because of  this reason, our GA+nKRR method is essentially as expensive as bare DFT. 
%
%%%For the systems considered in this work, about 97\% of the computational time of LDA+GA+ED calculations was used for solving recursively the embedding problem, while the rest was mainly utilized for DFT operations, such as constructing the Kohn-Sham Hamiltonian and calculating the electron density at each charge iteration. However, for LDA+GA+nKRR only about 45\% of the computational time is spent recursively solving the embedding problem with the rest used for the same DFT operations. 
%
In particular, the cost in terms of computational time and RAM is mainly determined by the EH solver within the GA+ED framework.
Instead, within GA+nKRR the computational bottleneck is determined by the DFT operations, such as constructing the Kohn-Sham Hamiltonian and calculating the electron density at each iteration (see the supplemental material).
For the calculations performed in this work, where the DFT part was performed using the all-electron scheme implemented in WIEN2k~\cite{WIEN2k}, utilizing the nKRR method for solving the EH equations reduced the computational time by a factor of 10 to 100 (depending on the system).
The relative computational gain of applying our methodology would presumably be even higher by utilizing less computationally-demanding implementations of DFT.

\section{Conclusions}

In summary, in this work we proposed a new computational framework
for simulating strongly-correlated electron systems, which offers the unique possibility of stepping up substantially the accuracy with respect to mean-field theories (such as classic approximations to DFT and DFT+U~\cite{LDA+U}), at a comparable computational cost.
This was accomplished combining QE theoretical frameworks with a fit-for-purpose ML technique (the nKRR), where the learning problem is facilitated by the $n$-mode expansion.
The fact that our method reduces the complexity of the learning problem from exponential to polynomial makes it realistically possible to 
extend it to systems with arbitrary stoichiometry and crystal structures.
Realizing this program will pave the way to virtually infinite applications in condensed matter physics, chemistry and materials science.

\section*{Methods}

All details relevant for reproducing our numerical results
are reported in the supplementary information of this manuscript.

\section*{Data availability statement}

Data are available from the authors upon reasonable request.

\section*{Acknowledgements}
We thank Gabriel Kotliar for useful discussions, Niels Carl W. Hansen for providing technical support
inherent in the computations performed at the Centre for Scientific Computing Aarhus (CSCAA) and Siraprapa Saraihom for editing Figs.~1,2.
We gratefully acknowledge funding from VILLUM FONDEN through the Villum Experiment project 00028019 and the Centre of Excellence for Dirac Materials (Grant. No. 11744). We also thank support from the Novo Nordisk Foundation through the Exploratory Interdisciplinary Synergy Programme project NNF19OC0057790.
T.-H.L. and Y.Y. were supported by the U.S. Department of Energy, Office of Science, Basic Energy Sciences,
as part of the Computational Materials Science Program.
T.-H.L. was also supported by the NSF Grant No. DMR-1733071.
V.D. was supported by NSF Grant No. 1822258.

\section*{Author contributions}
J.R. constructed and trained the nKRR algorithms.
J.R., T.-H.L., S.P. and W.X. performed the benchmark calculations of the actinides systems. 
V.D. co-supervised  J.R. and T.-H.L.. 
Y.Y. implemented the nKRR codes within the GA software CyGutz and provided technical support.
N.L. and O.C. conceived and led the project. All the authors contributed to the analysis and the interpretation of the results and to writing the manuscript.

%%%\bibliography{ref}

\begin{thebibliography}{70}%
\makeatletter
\providecommand \@ifxundefined [1]{%
 \@ifx{#1\undefined}
}%
\providecommand \@ifnum [1]{%
 \ifnum #1\expandafter \@firstoftwo
 \else \expandafter \@secondoftwo
 \fi
}%
\providecommand \@ifx [1]{%
 \ifx #1\expandafter \@firstoftwo
 \else \expandafter \@secondoftwo
 \fi
}%
\providecommand \natexlab [1]{#1}%
\providecommand \enquote  [1]{``#1''}%
\providecommand \bibnamefont  [1]{#1}%
\providecommand \bibfnamefont [1]{#1}%
\providecommand \citenamefont [1]{#1}%
\providecommand \href@noop [0]{\@secondoftwo}%
\providecommand \href [0]{\begingroup \@sanitize@url \@href}%
\providecommand \@href[1]{\@@startlink{#1}\@@href}%
\providecommand \@@href[1]{\endgroup#1\@@endlink}%
\providecommand \@sanitize@url [0]{\catcode `\\12\catcode `\$12\catcode
  `\&12\catcode `\#12\catcode `\^12\catcode `\_12\catcode `\%12\relax}%
\providecommand \@@startlink[1]{}%
\providecommand \@@endlink[0]{}%
\providecommand \url  [0]{\begingroup\@sanitize@url \@url }%
\providecommand \@url [1]{\endgroup\@href {#1}{\urlprefix }}%
\providecommand \urlprefix  [0]{URL }%
\providecommand \Eprint [0]{\href }%
\providecommand \doibase [0]{http://dx.doi.org/}%
\providecommand \selectlanguage [0]{\@gobble}%
\providecommand \bibinfo  [0]{\@secondoftwo}%
\providecommand \bibfield  [0]{\@secondoftwo}%
\providecommand \translation [1]{[#1]}%
\providecommand \BibitemOpen [0]{}%
\providecommand \bibitemStop [0]{}%
\providecommand \bibitemNoStop [0]{.\EOS\space}%
\providecommand \EOS [0]{\spacefactor3000\relax}%
\providecommand \BibitemShut  [1]{\csname bibitem#1\endcsname}%
\let\auto@bib@innerbib\@empty
%</preamble>
\bibitem [{\citenamefont {Mott}(1990)}]{Mott-book}%
  \BibitemOpen
  \bibfield  {author} {\bibinfo {author} {\bibfnamefont {N.~F}\ \bibnamefont
  {Mott}},\ }\href@noop {} {\emph {\bibinfo {title} {Metal-Insulator
  Transitions}}}\ (\bibinfo  {publisher} {Taylor and Francis,
  London/Philadelphia},\ \bibinfo {year} {1990})\BibitemShut {NoStop}%
\bibitem [{\citenamefont {Kent}\ and\ \citenamefont
  {Kotliar}(2018)}]{Kotliar-Science}%
  \BibitemOpen
  \bibfield  {author} {\bibinfo {author} {\bibfnamefont {Paul R.~C.}\
  \bibnamefont {Kent}}\ and\ \bibinfo {author} {\bibfnamefont {Gabriel}\
  \bibnamefont {Kotliar}},\ }\bibfield  {title} {\enquote {\bibinfo {title}
  {Toward a predictive theory of correlated materials},}\ }\href {\doibase
  10.1126/science.aat5975} {\bibfield  {journal} {\bibinfo  {journal}
  {Science}\ }\textbf {\bibinfo {volume} {361}},\ \bibinfo {pages} {348--354}
  (\bibinfo {year} {2018})}\BibitemShut {NoStop}%
\bibitem [{\citenamefont {Adler}\ \emph {et~al.}(2018)\citenamefont {Adler},
  \citenamefont {Kang}, \citenamefont {Yee},\ and\ \citenamefont
  {Kotliar}}]{Adler_2018}%
  \BibitemOpen
  \bibfield  {author} {\bibinfo {author} {\bibfnamefont {Ran}\ \bibnamefont
  {Adler}}, \bibinfo {author} {\bibfnamefont {Chang-Jong}\ \bibnamefont
  {Kang}}, \bibinfo {author} {\bibfnamefont {Chuck-Hou}\ \bibnamefont {Yee}}, \
  and\ \bibinfo {author} {\bibfnamefont {Gabriel}\ \bibnamefont {Kotliar}},\
  }\bibfield  {title} {\enquote {\bibinfo {title} {Correlated materials design:
  prospects and challenges},}\ }\href@noop {} {\bibfield  {journal} {\bibinfo
  {journal} {Reports on Progress in Physics}\ }\textbf {\bibinfo {volume}
  {82}},\ \bibinfo {pages} {012504} (\bibinfo {year} {2018})}\BibitemShut
  {NoStop}%
\bibitem [{\citenamefont {Lee}\ \emph {et~al.}(2006)\citenamefont {Lee},
  \citenamefont {Nagaosa},\ and\ \citenamefont {Wen}}]{highTc-1}%
  \BibitemOpen
  \bibfield  {author} {\bibinfo {author} {\bibfnamefont {Patrick~A.}\
  \bibnamefont {Lee}}, \bibinfo {author} {\bibfnamefont {Naoto}\ \bibnamefont
  {Nagaosa}}, \ and\ \bibinfo {author} {\bibfnamefont {Xiao-Gang}\ \bibnamefont
  {Wen}},\ }\bibfield  {title} {\enquote {\bibinfo {title} {Doping a mott
  insulator: Physics of high-temperature superconductivity},}\ }\href {\doibase
  10.1103/RevModPhys.78.17} {\bibfield  {journal} {\bibinfo  {journal} {Rev.
  Mod. Phys.}\ }\textbf {\bibinfo {volume} {78}},\ \bibinfo {pages} {17--85}
  (\bibinfo {year} {2006})}\BibitemShut {NoStop}%
\bibitem [{\citenamefont {Schiller}\ \emph {et~al.}(2015)\citenamefont
  {Schiller}, \citenamefont {Wagner},\ and\ \citenamefont
  {Ertekin}}]{corr-structure-1}%
  \BibitemOpen
  \bibfield  {author} {\bibinfo {author} {\bibfnamefont {Joshua~A.}\
  \bibnamefont {Schiller}}, \bibinfo {author} {\bibfnamefont {Lucas~K.}\
  \bibnamefont {Wagner}}, \ and\ \bibinfo {author} {\bibfnamefont {Elif}\
  \bibnamefont {Ertekin}},\ }\bibfield  {title} {\enquote {\bibinfo {title}
  {Phase stability and properties of manganese oxide polymorphs: Assessment and
  insights from diffusion $\text{Monte Carlo}$},}\ }\href@noop {} {\bibfield
  {journal} {\bibinfo  {journal} {Phys. Rev. B}\ }\textbf {\bibinfo {volume}
  {92}},\ \bibinfo {pages} {235209} (\bibinfo {year} {2015})}\BibitemShut
  {NoStop}%
\bibitem [{\citenamefont {Schr\"on}\ \emph {et~al.}(2010)\citenamefont
  {Schr\"on}, \citenamefont {R\"odl},\ and\ \citenamefont
  {Bechstedt}}]{corr-structure-3}%
  \BibitemOpen
  \bibfield  {author} {\bibinfo {author} {\bibfnamefont {A.}~\bibnamefont
  {Schr\"on}}, \bibinfo {author} {\bibfnamefont {C.}~\bibnamefont {R\"odl}}, \
  and\ \bibinfo {author} {\bibfnamefont {F.}~\bibnamefont {Bechstedt}},\
  }\bibfield  {title} {\enquote {\bibinfo {title} {Energetic stability and
  magnetic properties of $\text{MnO}$ in the rocksalt, wurtzite, and
  zinc-blende structures: Influence of exchange and correlation},}\ }\href@noop
  {} {\bibfield  {journal} {\bibinfo  {journal} {Phys. Rev. B}\ }\textbf
  {\bibinfo {volume} {82}},\ \bibinfo {pages} {165109} (\bibinfo {year}
  {2010})}\BibitemShut {NoStop}%
\bibitem [{\citenamefont {Koloren\ifmmode~\check{c}\else \v{c}\fi{}}\ and\
  \citenamefont {Mitas}(2008)}]{corr-structure-4}%
  \BibitemOpen
  \bibfield  {author} {\bibinfo {author} {\bibfnamefont {Jind\ifmmode
  \check{r}\else~\v{r}\fi{}ich}\ \bibnamefont {Koloren\ifmmode~\check{c}\else
  \v{c}\fi{}}}\ and\ \bibinfo {author} {\bibfnamefont {Lubos}\ \bibnamefont
  {Mitas}},\ }\bibfield  {title} {\enquote {\bibinfo {title} {Quantum
  $\text{Monte Carlo}$ calculations of structural properties of $\text{FeO}$
  under pressure},}\ }\href@noop {} {\bibfield  {journal} {\bibinfo  {journal}
  {Phys. Rev. Lett.}\ }\textbf {\bibinfo {volume} {101}},\ \bibinfo {pages}
  {185502} (\bibinfo {year} {2008})}\BibitemShut {NoStop}%
\bibitem [{\citenamefont {Leonov}\ \emph {et~al.}(2016)\citenamefont {Leonov},
  \citenamefont {Pourovskii}, \citenamefont {Georges},\ and\ \citenamefont
  {Abrikosov}}]{corr-structure-5}%
  \BibitemOpen
  \bibfield  {author} {\bibinfo {author} {\bibfnamefont {I.}~\bibnamefont
  {Leonov}}, \bibinfo {author} {\bibfnamefont {L.}~\bibnamefont {Pourovskii}},
  \bibinfo {author} {\bibfnamefont {A.}~\bibnamefont {Georges}}, \ and\
  \bibinfo {author} {\bibfnamefont {I.~A.}\ \bibnamefont {Abrikosov}},\
  }\bibfield  {title} {\enquote {\bibinfo {title} {Magnetic collapse and the
  behavior of transition metal oxides at high pressure},}\ }\href@noop {}
  {\bibfield  {journal} {\bibinfo  {journal} {Phys. Rev. B}\ }\textbf {\bibinfo
  {volume} {94}},\ \bibinfo {pages} {155135} (\bibinfo {year}
  {2016})}\BibitemShut {NoStop}%
\bibitem [{\citenamefont {Leonov}\ \emph {et~al.}(2014)\citenamefont {Leonov},
  \citenamefont {Anisimov},\ and\ \citenamefont {Vollhardt}}]{dmft_forces}%
  \BibitemOpen
  \bibfield  {author} {\bibinfo {author} {\bibfnamefont {I.}~\bibnamefont
  {Leonov}}, \bibinfo {author} {\bibfnamefont {V.~I.}\ \bibnamefont
  {Anisimov}}, \ and\ \bibinfo {author} {\bibfnamefont {D.}~\bibnamefont
  {Vollhardt}},\ }\bibfield  {title} {\enquote {\bibinfo {title}
  {First-principles calculation of atomic forces and structural distortions in
  strongly correlated materials},}\ }\href@noop {} {\bibfield  {journal}
  {\bibinfo  {journal} {Phys. Rev. Lett.}\ }\textbf {\bibinfo {volume} {112}},\
  \bibinfo {pages} {146401} (\bibinfo {year} {2014})}\BibitemShut {NoStop}%
\bibitem [{\citenamefont {Lanat\`a}\ \emph {et~al.}(2019)\citenamefont
  {Lanat\`a}, \citenamefont {Tsung-Han}, \citenamefont {Yong-Xin},\ and\
  \citenamefont {Dobrosavljevi\'c}}]{npj-lanata}%
  \BibitemOpen
  \bibfield  {author} {\bibinfo {author} {\bibfnamefont {Nicola}\ \bibnamefont
  {Lanat\`a}}, \bibinfo {author} {\bibfnamefont {Lee}\ \bibnamefont
  {Tsung-Han}}, \bibinfo {author} {\bibfnamefont {Vladan~Stevanovi\'c}\
  \bibnamefont {Yong-Xin}, \bibfnamefont {Yao}}, \ and\ \bibinfo {author}
  {\bibfnamefont {Vladimir}\ \bibnamefont {Dobrosavljevi\'c}},\ }\bibfield
  {title} {\enquote {\bibinfo {title} {Connection between mott physics and
  crystal structure in a series of transition metal binary compounds},}\
  }\href@noop {} {\bibfield  {journal} {\bibinfo  {journal} {npj Comput.
  Mater.}\ }\textbf {\bibinfo {volume} {5}},\ \bibinfo {pages} {30} (\bibinfo
  {year} {2019})}\BibitemShut {NoStop}%
\bibitem [{\citenamefont {Hohenberg}\ and\ \citenamefont
  {Kohn}(1964)}]{HohenbergandKohn}%
  \BibitemOpen
  \bibfield  {author} {\bibinfo {author} {\bibfnamefont {P.}~\bibnamefont
  {Hohenberg}}\ and\ \bibinfo {author} {\bibfnamefont {W.}~\bibnamefont
  {Kohn}},\ }\bibfield  {title} {\enquote {\bibinfo {title} {Inhomogeneous
  electron gas},}\ }\href {\doibase 10.1103/PhysRev.136.B864} {\bibfield
  {journal} {\bibinfo  {journal} {Phys. Rev.}\ }\textbf {\bibinfo {volume}
  {136}},\ \bibinfo {pages} {B864} (\bibinfo {year} {1964})}\BibitemShut
  {NoStop}%
\bibitem [{\citenamefont {Kohn}\ and\ \citenamefont
  {Sham}(1965)}]{KohnandSham}%
  \BibitemOpen
  \bibfield  {author} {\bibinfo {author} {\bibfnamefont {W.}~\bibnamefont
  {Kohn}}\ and\ \bibinfo {author} {\bibfnamefont {L.~J.}\ \bibnamefont
  {Sham}},\ }\bibfield  {title} {\enquote {\bibinfo {title} {Self-consistent
  equations including exchange and correlation effects},}\ }\href {\doibase
  10.1103/PhysRev.140.A1133} {\bibfield  {journal} {\bibinfo  {journal} {Phys.
  Rev.}\ }\textbf {\bibinfo {volume} {140}},\ \bibinfo {pages} {A1133}
  (\bibinfo {year} {1965})}\BibitemShut {NoStop}%
\bibitem [{\citenamefont {Gunnarsson}\ and\ \citenamefont
  {Lundqvist}(1976)}]{LDA}%
  \BibitemOpen
  \bibfield  {author} {\bibinfo {author} {\bibfnamefont {O.}~\bibnamefont
  {Gunnarsson}}\ and\ \bibinfo {author} {\bibfnamefont {B.~I.}\ \bibnamefont
  {Lundqvist}},\ }\bibfield  {title} {\enquote {\bibinfo {title} {Exchange and
  correlation in atoms, molecules, and solids by the spin-density-functional
  formalism},}\ }\href {\doibase 10.1103/PhysRevB.13.4274} {\bibfield
  {journal} {\bibinfo  {journal} {Phys. Rev. B}\ }\textbf {\bibinfo {volume}
  {13}},\ \bibinfo {pages} {4274} (\bibinfo {year} {1976})}\BibitemShut
  {NoStop}%
\bibitem [{\citenamefont {Perdew}\ \emph {et~al.}(1996)\citenamefont {Perdew},
  \citenamefont {Burke},\ and\ \citenamefont {Ernzerhof}}]{GGA-simple}%
  \BibitemOpen
  \bibfield  {author} {\bibinfo {author} {\bibfnamefont {John~P.}\ \bibnamefont
  {Perdew}}, \bibinfo {author} {\bibfnamefont {Kieron}\ \bibnamefont {Burke}},
  \ and\ \bibinfo {author} {\bibfnamefont {Matthias}\ \bibnamefont
  {Ernzerhof}},\ }\bibfield  {title} {\enquote {\bibinfo {title} {Generalized
  gradient approximation made simple},}\ }\href {\doibase
  10.1103/PhysRevLett.77.3865} {\bibfield  {journal} {\bibinfo  {journal}
  {Phys. Rev. Lett.}\ }\textbf {\bibinfo {volume} {77}},\ \bibinfo {pages}
  {3865--3868} (\bibinfo {year} {1996})}\BibitemShut {NoStop}%
\bibitem [{\citenamefont {Jones}(2015)}]{DFT-RevModPhys.87.897}%
  \BibitemOpen
  \bibfield  {author} {\bibinfo {author} {\bibfnamefont {R.~O.}\ \bibnamefont
  {Jones}},\ }\bibfield  {title} {\enquote {\bibinfo {title} {Density
  functional theory: Its origins, rise to prominence, and future},}\ }\href
  {\doibase 10.1103/RevModPhys.87.897} {\bibfield  {journal} {\bibinfo
  {journal} {Rev. Mod. Phys.}\ }\textbf {\bibinfo {volume} {87}},\ \bibinfo
  {pages} {897--923} (\bibinfo {year} {2015})}\BibitemShut {NoStop}%
\bibitem [{\citenamefont {Mardirossian}\ and\ \citenamefont
  {Head-Gordon}(2017)}]{DFT-doi:10.1080/00268976.2017.1333644}%
  \BibitemOpen
  \bibfield  {author} {\bibinfo {author} {\bibfnamefont {Narbe}\ \bibnamefont
  {Mardirossian}}\ and\ \bibinfo {author} {\bibfnamefont {Martin}\ \bibnamefont
  {Head-Gordon}},\ }\bibfield  {title} {\enquote {\bibinfo {title} {Thirty
  years of density functional theory in computational chemistry: an overview
  and extensive assessment of 200 density functionals},}\ }\href@noop {}
  {\bibfield  {journal} {\bibinfo  {journal} {Molecular Physics}\ }\textbf
  {\bibinfo {volume} {115}},\ \bibinfo {pages} {2315--2372} (\bibinfo {year}
  {2017})}\BibitemShut {NoStop}%
\bibitem [{\citenamefont {Burke}(2012)}]{DFT-doi:10.1063/1.4704546}%
  \BibitemOpen
  \bibfield  {author} {\bibinfo {author} {\bibfnamefont {Kieron}\ \bibnamefont
  {Burke}},\ }\bibfield  {title} {\enquote {\bibinfo {title} {Perspective on
  density functional theory},}\ }\href {\doibase 10.1063/1.4704546} {\bibfield
  {journal} {\bibinfo  {journal} {The Journal of Chemical Physics}\ }\textbf
  {\bibinfo {volume} {136}},\ \bibinfo {pages} {150901} (\bibinfo {year}
  {2012})}\BibitemShut {NoStop}%
\bibitem [{\citenamefont {Qiming}\ and\ \citenamefont
  {Garnet~Kin-Lic}(2016)}]{quantum-embedding-review}%
  \BibitemOpen
  \bibfield  {author} {\bibinfo {author} {\bibfnamefont {Sun}\ \bibnamefont
  {Qiming}}\ and\ \bibinfo {author} {\bibfnamefont {Chan}\ \bibnamefont
  {Garnet~Kin-Lic}},\ }\bibfield  {title} {\enquote {\bibinfo {title} {Quantum
  embedding theories},}\ }\href@noop {} {\bibfield  {journal} {\bibinfo
  {journal} {Acc. Chem. Res.}\ }\textbf {\bibinfo {volume} {49}},\ \bibinfo
  {pages} {2705} (\bibinfo {year} {2016})}\BibitemShut {NoStop}%
\bibitem [{\citenamefont {Fulde}\ and\ \citenamefont
  {Stoll}(2017)}]{quantum-embedding-another}%
  \BibitemOpen
  \bibfield  {author} {\bibinfo {author} {\bibfnamefont {Peter}\ \bibnamefont
  {Fulde}}\ and\ \bibinfo {author} {\bibfnamefont {Hermann}\ \bibnamefont
  {Stoll}},\ }\bibfield  {title} {\enquote {\bibinfo {title} {Dealing with the
  exponential wall in electronic structure calculations},}\ }\href@noop {}
  {\bibfield  {journal} {\bibinfo  {journal} {The Journal of Chemical Physics}\
  }\textbf {\bibinfo {volume} {146}},\ \bibinfo {pages} {194107} (\bibinfo
  {year} {2017})}\BibitemShut {NoStop}%
\bibitem [{\citenamefont {Georges}\ \emph {et~al.}(1996)\citenamefont
  {Georges}, \citenamefont {Kotliar}, \citenamefont {Krauth},\ and\
  \citenamefont {Rozenberg}}]{DMFT}%
  \BibitemOpen
  \bibfield  {author} {\bibinfo {author} {\bibfnamefont {A.}~\bibnamefont
  {Georges}}, \bibinfo {author} {\bibfnamefont {G.}~\bibnamefont {Kotliar}},
  \bibinfo {author} {\bibfnamefont {W.}~\bibnamefont {Krauth}}, \ and\ \bibinfo
  {author} {\bibfnamefont {M.~J.}\ \bibnamefont {Rozenberg}},\ }\bibfield
  {title} {\enquote {\bibinfo {title} {Dynamical mean-field theory of strongly
  correlated fermion systems and the limit of infinite dimensions},}\
  }\href@noop {} {\bibfield  {journal} {\bibinfo  {journal} {Rev. Mod. Phys.}\
  }\textbf {\bibinfo {volume} {68}},\ \bibinfo {pages} {13} (\bibinfo {year}
  {1996})}\BibitemShut {NoStop}%
\bibitem [{\citenamefont {Anisimov}\ and\ \citenamefont
  {Izyumov}(2010)}]{dmft_book}%
  \BibitemOpen
  \bibfield  {author} {\bibinfo {author} {\bibfnamefont {V.}~\bibnamefont
  {Anisimov}}\ and\ \bibinfo {author} {\bibfnamefont {Y.}~\bibnamefont
  {Izyumov}},\ }\href@noop {} {\emph {\bibinfo {title} {Electronic Structure of
  Strongly Correlated Materials}}}\ (\bibinfo  {publisher} {Springer},\
  \bibinfo {year} {2010})\BibitemShut {NoStop}%
\bibitem [{\citenamefont {Kotliar}\ \emph {et~al.}(2006)\citenamefont
  {Kotliar}, \citenamefont {Savrasov}, \citenamefont {Haule}, \citenamefont
  {Oudovenko}, \citenamefont {Parcollet},\ and\ \citenamefont
  {Marianetti}}]{LDA+U+DMFT}%
  \BibitemOpen
  \bibfield  {author} {\bibinfo {author} {\bibfnamefont {G.}~\bibnamefont
  {Kotliar}}, \bibinfo {author} {\bibfnamefont {S.~Y.}\ \bibnamefont
  {Savrasov}}, \bibinfo {author} {\bibfnamefont {K.}~\bibnamefont {Haule}},
  \bibinfo {author} {\bibfnamefont {V.~S.}\ \bibnamefont {Oudovenko}}, \bibinfo
  {author} {\bibfnamefont {O.}~\bibnamefont {Parcollet}}, \ and\ \bibinfo
  {author} {\bibfnamefont {C.~A.}\ \bibnamefont {Marianetti}},\ }\bibfield
  {title} {\enquote {\bibinfo {title} {Electronic structure calculations with
  dynamical mean-field theory},}\ }\href@noop {} {\bibfield  {journal}
  {\bibinfo  {journal} {Rev. Mod. Phys.}\ }\textbf {\bibinfo {volume} {78}},\
  \bibinfo {eid} {865} (\bibinfo {year} {2006})}\BibitemShut {NoStop}%
\bibitem [{\citenamefont {Held}\ \emph {et~al.}(2006)\citenamefont {Held},
  \citenamefont {Nekrasov}, \citenamefont {Keller}, \citenamefont {Eyert},
  \citenamefont {Bl\"umer}, \citenamefont {McMahan}, \citenamefont {Scalettar},
  \citenamefont {Pruschke}, \citenamefont {Anisimov},\ and\ \citenamefont
  {Vollhardt}}]{Held-review-DMFT}%
  \BibitemOpen
  \bibfield  {author} {\bibinfo {author} {\bibfnamefont {K.}~\bibnamefont
  {Held}}, \bibinfo {author} {\bibfnamefont {A.}~\bibnamefont {Nekrasov}},
  \bibinfo {author} {\bibfnamefont {G.}~\bibnamefont {Keller}}, \bibinfo
  {author} {\bibfnamefont {V.}~\bibnamefont {Eyert}}, \bibinfo {author}
  {\bibfnamefont {N.}~\bibnamefont {Bl\"umer}}, \bibinfo {author}
  {\bibfnamefont {A.~K.}\ \bibnamefont {McMahan}}, \bibinfo {author}
  {\bibfnamefont {R.~T.}\ \bibnamefont {Scalettar}}, \bibinfo {author}
  {\bibfnamefont {Th.}\ \bibnamefont {Pruschke}}, \bibinfo {author}
  {\bibfnamefont {V.~I.}\ \bibnamefont {Anisimov}}, \ and\ \bibinfo {author}
  {\bibfnamefont {D.}~\bibnamefont {Vollhardt}},\ }\bibfield  {title} {\enquote
  {\bibinfo {title} {Realistic investigations of correlated electron systems
  with $\text{LDA+DMFT}$},}\ }\href@noop {} {\bibfield  {journal} {\bibinfo
  {journal} {Phys. Stat. Sol. (B)}\ }\textbf {\bibinfo {volume} {243}},\
  \bibinfo {pages} {2599} (\bibinfo {year} {2006})}\BibitemShut {NoStop}%
\bibitem [{\citenamefont {Anisimov}\ \emph
  {et~al.}(1997{\natexlab{a}})\citenamefont {Anisimov}, \citenamefont
  {Oteryaev}, \citenamefont {Korotin}, \citenamefont {Anokhin},\ and\
  \citenamefont {Kotliar}}]{Anisimov_DMFT}%
  \BibitemOpen
  \bibfield  {author} {\bibinfo {author} {\bibfnamefont {V.~I.}\ \bibnamefont
  {Anisimov}}, \bibinfo {author} {\bibfnamefont {A.~I.}\ \bibnamefont
  {Oteryaev}}, \bibinfo {author} {\bibfnamefont {M.~A.}\ \bibnamefont
  {Korotin}}, \bibinfo {author} {\bibfnamefont {A.~O.}\ \bibnamefont
  {Anokhin}}, \ and\ \bibinfo {author} {\bibfnamefont {G.}~\bibnamefont
  {Kotliar}},\ }\bibfield  {title} {\enquote {\bibinfo {title}
  {First-principles calculations of the electronic structure and spectra of
  strongly correlated systems: dynamical mean-field theory},}\ }\href@noop {}
  {\bibfield  {journal} {\bibinfo  {journal} {J. Phys. Condens. Matter}\
  }\textbf {\bibinfo {volume} {9}},\ \bibinfo {pages} {7359} (\bibinfo {year}
  {1997}{\natexlab{a}})}\BibitemShut {NoStop}%
\bibitem [{\citenamefont {Maier}\ \emph {et~al.}(2005)\citenamefont {Maier},
  \citenamefont {Jarrell}, \citenamefont {Pruschke},\ and\ \citenamefont
  {Hettler}}]{CDMFT-Jarrell}%
  \BibitemOpen
  \bibfield  {author} {\bibinfo {author} {\bibfnamefont {T.}~\bibnamefont
  {Maier}}, \bibinfo {author} {\bibfnamefont {M.}~\bibnamefont {Jarrell}},
  \bibinfo {author} {\bibfnamefont {T.}~\bibnamefont {Pruschke}}, \ and\
  \bibinfo {author} {\bibfnamefont {M.~H.}\ \bibnamefont {Hettler}},\
  }\bibfield  {title} {\enquote {\bibinfo {title} {Quantum cluster theories},}\
  }\href@noop {} {\bibfield  {journal} {\bibinfo  {journal} {Rev. Mod. Phys.}\
  }\textbf {\bibinfo {volume} {77}},\ \bibinfo {pages} {1027} (\bibinfo {year}
  {2005})}\BibitemShut {NoStop}%
\bibitem [{\citenamefont {Potthoff}\ \emph {et~al.}(2003)\citenamefont
  {Potthoff}, \citenamefont {Aichhorn},\ and\ \citenamefont
  {Dahnken}}]{CDMFT-Potthoff}%
  \BibitemOpen
  \bibfield  {author} {\bibinfo {author} {\bibfnamefont {M.}~\bibnamefont
  {Potthoff}}, \bibinfo {author} {\bibfnamefont {M.}~\bibnamefont {Aichhorn}},
  \ and\ \bibinfo {author} {\bibfnamefont {C.}~\bibnamefont {Dahnken}},\
  }\bibfield  {title} {\enquote {\bibinfo {title} {Variational cluster approach
  to correlated electron systems in low dimensions},}\ }\href@noop {}
  {\bibfield  {journal} {\bibinfo  {journal} {Phys. Rev. Lett.}\ }\textbf
  {\bibinfo {volume} {91}},\ \bibinfo {pages} {206402} (\bibinfo {year}
  {2003})}\BibitemShut {NoStop}%
\bibitem [{\citenamefont {Lichtenstein}\ and\ \citenamefont
  {Katsnelson}(2000)}]{CDMFT-Lichtenstein}%
  \BibitemOpen
  \bibfield  {author} {\bibinfo {author} {\bibfnamefont {A.~I.}\ \bibnamefont
  {Lichtenstein}}\ and\ \bibinfo {author} {\bibfnamefont {M.~I.}\ \bibnamefont
  {Katsnelson}},\ }\bibfield  {title} {\enquote {\bibinfo {title}
  {Antiferromagnetism and d-wave superconductivity in cuprates: A cluster
  dynamical mean-field theory},}\ }\href@noop {} {\bibfield  {journal}
  {\bibinfo  {journal} {Phys. Rev. B}\ }\textbf {\bibinfo {volume} {62}},\
  \bibinfo {pages} {R9283} (\bibinfo {year} {2000})}\BibitemShut {NoStop}%
\bibitem [{\citenamefont {Knizia}\ and\ \citenamefont {Chan}(2012)}]{DMET}%
  \BibitemOpen
  \bibfield  {author} {\bibinfo {author} {\bibfnamefont {G.}~\bibnamefont
  {Knizia}}\ and\ \bibinfo {author} {\bibfnamefont {G.~K.-L.}\ \bibnamefont
  {Chan}},\ }\bibfield  {title} {\enquote {\bibinfo {title} {Density matrix
  embedding: A simple alternative to dynamical mean-field theory},}\
  }\href@noop {} {\bibfield  {journal} {\bibinfo  {journal} {Phys. Rev. Lett.}\
  }\textbf {\bibinfo {volume} {109}},\ \bibinfo {pages} {186404} (\bibinfo
  {year} {2012})}\BibitemShut {NoStop}%
\bibitem [{\citenamefont {Bulik}\ \emph {et~al.}(2014)\citenamefont {Bulik},
  \citenamefont {Scuseria},\ and\ \citenamefont {Dukelsky}}]{Bulik-DMET}%
  \BibitemOpen
  \bibfield  {author} {\bibinfo {author} {\bibfnamefont {I.~W.}\ \bibnamefont
  {Bulik}}, \bibinfo {author} {\bibfnamefont {G.~E.}\ \bibnamefont {Scuseria}},
  \ and\ \bibinfo {author} {\bibfnamefont {J.}~\bibnamefont {Dukelsky}},\
  }\bibfield  {title} {\enquote {\bibinfo {title} {Density matrix embedding
  from broken symmetry lattice mean fields},}\ }\href@noop {} {\bibfield
  {journal} {\bibinfo  {journal} {Phys. Rev. B}\ }\textbf {\bibinfo {volume}
  {89}},\ \bibinfo {pages} {035140} (\bibinfo {year} {2014})}\BibitemShut
  {NoStop}%
\bibitem [{\citenamefont {Lanat\`a}\ \emph {et~al.}(2015)\citenamefont
  {Lanat\`a}, \citenamefont {Yao}, \citenamefont {Wang}, \citenamefont {Ho},\
  and\ \citenamefont {Kotliar}}]{Our-PRX}%
  \BibitemOpen
  \bibfield  {author} {\bibinfo {author} {\bibfnamefont {Nicola}\ \bibnamefont
  {Lanat\`a}}, \bibinfo {author} {\bibfnamefont {Yong~Xin}\ \bibnamefont
  {Yao}}, \bibinfo {author} {\bibfnamefont {Cai-Zhuang}\ \bibnamefont {Wang}},
  \bibinfo {author} {\bibfnamefont {Kai-Ming}\ \bibnamefont {Ho}}, \ and\
  \bibinfo {author} {\bibfnamefont {Gabriel}\ \bibnamefont {Kotliar}},\
  }\bibfield  {title} {\enquote {\bibinfo {title} {Phase diagram and electronic
  structure of praseodymium and plutonium},}\ }\href@noop {} {\bibfield
  {journal} {\bibinfo  {journal} {Phys. Rev. X}\ }\textbf {\bibinfo {volume}
  {5}},\ \bibinfo {pages} {011008} (\bibinfo {year} {2015})}\BibitemShut
  {NoStop}%
\bibitem [{\citenamefont {Gutzwiller}(1965)}]{Gutzwiller3}%
  \BibitemOpen
  \bibfield  {author} {\bibinfo {author} {\bibfnamefont {M.~C.}\ \bibnamefont
  {Gutzwiller}},\ }\bibfield  {title} {\enquote {\bibinfo {title} {Correlation
  of {Electrons} in a {Narrow} {$s$} {Band}},}\ }\href {\doibase
  10.1103/PhysRev.137.A1726} {\bibfield  {journal} {\bibinfo  {journal} {Phys.
  Rev.}\ }\textbf {\bibinfo {volume} {137}},\ \bibinfo {pages} {A1726}
  (\bibinfo {year} {1965})}\BibitemShut {NoStop}%
\bibitem [{\citenamefont {B\"unemann}\ \emph {et~al.}(1998)\citenamefont
  {B\"unemann}, \citenamefont {Weber},\ and\ \citenamefont
  {Gebhard}}]{Gebhard}%
  \BibitemOpen
  \bibfield  {author} {\bibinfo {author} {\bibfnamefont {J.}~\bibnamefont
  {B\"unemann}}, \bibinfo {author} {\bibfnamefont {W.}~\bibnamefont {Weber}}, \
  and\ \bibinfo {author} {\bibfnamefont {F.}~\bibnamefont {Gebhard}},\
  }\bibfield  {title} {\enquote {\bibinfo {title} {Multiband
  $\text{Gutzwiller}$ wave functions for general on-site interactions},}\
  }\href@noop {} {\bibfield  {journal} {\bibinfo  {journal} {Phys. Rev. B}\
  }\textbf {\bibinfo {volume} {57}},\ \bibinfo {pages} {6896} (\bibinfo {year}
  {1998})}\BibitemShut {NoStop}%
\bibitem [{\citenamefont {Deng}\ \emph {et~al.}(2009)\citenamefont {Deng},
  \citenamefont {Wang}, \citenamefont {Dai},\ and\ \citenamefont
  {Fang}}]{Fang}%
  \BibitemOpen
  \bibfield  {author} {\bibinfo {author} {\bibfnamefont {X.-Y.}\ \bibnamefont
  {Deng}}, \bibinfo {author} {\bibfnamefont {L.}~\bibnamefont {Wang}}, \bibinfo
  {author} {\bibfnamefont {X.}~\bibnamefont {Dai}}, \ and\ \bibinfo {author}
  {\bibfnamefont {Z.}~\bibnamefont {Fang}},\ }\bibfield  {title} {\enquote
  {\bibinfo {title} {Local density approximation combined with {Gutzwiller}
  method for correlated electron systems: {Formalism} and applications},}\
  }\href@noop {} {\bibfield  {journal} {\bibinfo  {journal} {Phys. Rev. B}\
  }\textbf {\bibinfo {volume} {79}},\ \bibinfo {eid} {075114} (\bibinfo {year}
  {2009})}\BibitemShut {NoStop}%
\bibitem [{\citenamefont {Lanat\`a}\ \emph {et~al.}(2017)\citenamefont
  {Lanat\`a}, \citenamefont {Lee}, \citenamefont {Yao},\ and\ \citenamefont
  {Dobrosavljevi\ifmmode~\acute{c}\else \'{c}\fi{}}}]{Ghost-GA}%
  \BibitemOpen
  \bibfield  {author} {\bibinfo {author} {\bibfnamefont {Nicola}\ \bibnamefont
  {Lanat\`a}}, \bibinfo {author} {\bibfnamefont {Tsung-Han}\ \bibnamefont
  {Lee}}, \bibinfo {author} {\bibfnamefont {Yong-Xin}\ \bibnamefont {Yao}}, \
  and\ \bibinfo {author} {\bibfnamefont {Vladimir}\ \bibnamefont
  {Dobrosavljevi\ifmmode~\acute{c}\else \'{c}\fi{}}},\ }\bibfield  {title}
  {\enquote {\bibinfo {title} {Emergent bloch excitations in mott matter},}\
  }\href@noop {} {\bibfield  {journal} {\bibinfo  {journal} {Phys. Rev. B}\
  }\textbf {\bibinfo {volume} {96}},\ \bibinfo {pages} {195126} (\bibinfo
  {year} {2017})}\BibitemShut {NoStop}%
\bibitem [{\citenamefont {Fr{\'{e}}sard}\ and\ \citenamefont
  {W{\"{o}}lfle}(1992)}]{Fresard1992}%
  \BibitemOpen
  \bibfield  {author} {\bibinfo {author} {\bibfnamefont {R.}~\bibnamefont
  {Fr{\'{e}}sard}}\ and\ \bibinfo {author} {\bibfnamefont {P.}~\bibnamefont
  {W{\"{o}}lfle}},\ }\bibfield  {title} {\enquote {\bibinfo {title} {{Unified
  Slave Boson Representation of Spin and Charge Degrees of Freedom for Strongly
  Correlated Fermi Systems}},}\ }\href {\doibase 10.1142/S0217979292000414}
  {\bibfield  {journal} {\bibinfo  {journal} {International Journal of Modern
  Physics B}\ }\textbf {\bibinfo {volume} {06}},\ \bibinfo {pages} {685--704}
  (\bibinfo {year} {1992})}\BibitemShut {NoStop}%
\bibitem [{\citenamefont {Lechermann}\ \emph {et~al.}(2007)\citenamefont
  {Lechermann}, \citenamefont {Georges}, \citenamefont {Kotliar},\ and\
  \citenamefont {Parcollet}}]{Georges}%
  \BibitemOpen
  \bibfield  {author} {\bibinfo {author} {\bibfnamefont {F.}~\bibnamefont
  {Lechermann}}, \bibinfo {author} {\bibfnamefont {A.}~\bibnamefont {Georges}},
  \bibinfo {author} {\bibfnamefont {G.}~\bibnamefont {Kotliar}}, \ and\
  \bibinfo {author} {\bibfnamefont {O.}~\bibnamefont {Parcollet}},\ }\bibfield
  {title} {\enquote {\bibinfo {title} {Rotationally invariant slave-boson
  formalism and momentum dependence of the quasiparticle weight},}\ }\href@noop
  {} {\bibfield  {journal} {\bibinfo  {journal} {Phys. Rev. B}\ }\textbf
  {\bibinfo {volume} {76}},\ \bibinfo {eid} {155102} (\bibinfo {year}
  {2007})}\BibitemShut {NoStop}%
\bibitem [{\citenamefont {Lanat{\`{a}}}\ \emph {et~al.}(2017)\citenamefont
  {Lanat{\`{a}}}, \citenamefont {Yao}, \citenamefont {Deng}, \citenamefont
  {Dobrosavljevi{\'{c}}},\ and\ \citenamefont {Kotliar}}]{Lanata2016}%
  \BibitemOpen
  \bibfield  {author} {\bibinfo {author} {\bibfnamefont {Nicola}\ \bibnamefont
  {Lanat{\`{a}}}}, \bibinfo {author} {\bibfnamefont {Yongxin}\ \bibnamefont
  {Yao}}, \bibinfo {author} {\bibfnamefont {Xiaoyu}\ \bibnamefont {Deng}},
  \bibinfo {author} {\bibfnamefont {Vladimir}\ \bibnamefont
  {Dobrosavljevi{\'{c}}}}, \ and\ \bibinfo {author} {\bibfnamefont {Gabriel}\
  \bibnamefont {Kotliar}},\ }\bibfield  {title} {\enquote {\bibinfo {title}
  {{Slave Boson Theory of Orbital Differentiation with Crystal Field Effects:
  Application to UO$_2$}},}\ }\href {\doibase 10.1103/PhysRevLett.118.126401}
  {\bibfield  {journal} {\bibinfo  {journal} {Physical Review Letters}\
  }\textbf {\bibinfo {volume} {118}},\ \bibinfo {pages} {126401} (\bibinfo
  {year} {2017})}\BibitemShut {NoStop}%
\bibitem [{\citenamefont {B\"unemann}\ and\ \citenamefont
  {Gebhard}(2007)}]{equivalence_GA-SB}%
  \BibitemOpen
  \bibfield  {author} {\bibinfo {author} {\bibfnamefont {J.}~\bibnamefont
  {B\"unemann}}\ and\ \bibinfo {author} {\bibfnamefont {F.}~\bibnamefont
  {Gebhard}},\ }\bibfield  {title} {\enquote {\bibinfo {title} {Equivalence of
  $\text{Gutzwiller}$ and slave-boson mean-field theories for multiband
  $\text{Hubbard}$ models},}\ }\href@noop {} {\bibfield  {journal} {\bibinfo
  {journal} {Phys. Rev. B}\ }\textbf {\bibinfo {volume} {76}},\ \bibinfo
  {pages} {193104} (\bibinfo {year} {2007})}\BibitemShut {NoStop}%
\bibitem [{\citenamefont {Lanat\`{a}}\ \emph {et~al.}(2008)\citenamefont
  {Lanat\`{a}}, \citenamefont {Barone},\ and\ \citenamefont
  {Fabrizio}}]{lanata-barone-fabrizio}%
  \BibitemOpen
  \bibfield  {author} {\bibinfo {author} {\bibfnamefont {N.}~\bibnamefont
  {Lanat\`{a}}}, \bibinfo {author} {\bibfnamefont {P.}~\bibnamefont {Barone}},
  \ and\ \bibinfo {author} {\bibfnamefont {M.}~\bibnamefont {Fabrizio}},\
  }\bibfield  {title} {\enquote {\bibinfo {title} {Fermi-surface evolution
  across the magnetic phase transition in the {Kondo} lattice model},}\
  }\href@noop {} {\bibfield  {journal} {\bibinfo  {journal} {Phys. Rev. B}\
  }\textbf {\bibinfo {volume} {78}},\ \bibinfo {eid} {155127} (\bibinfo {year}
  {2008})}\BibitemShut {NoStop}%
\bibitem [{\citenamefont {Ayral}\ \emph {et~al.}(2017)\citenamefont {Ayral},
  \citenamefont {Lee},\ and\ \citenamefont {Kotliar}}]{dmet-risb-1}%
  \BibitemOpen
  \bibfield  {author} {\bibinfo {author} {\bibfnamefont {Thomas}\ \bibnamefont
  {Ayral}}, \bibinfo {author} {\bibfnamefont {Tsung-Han}\ \bibnamefont {Lee}},
  \ and\ \bibinfo {author} {\bibfnamefont {Gabriel}\ \bibnamefont {Kotliar}},\
  }\bibfield  {title} {\enquote {\bibinfo {title} {Dynamical mean-field theory,
  density-matrix embedding theory, and rotationally invariant slave bosons: A
  unified perspective},}\ }\href {\doibase 10.1103/PhysRevB.96.235139}
  {\bibfield  {journal} {\bibinfo  {journal} {Phys. Rev. B}\ }\textbf {\bibinfo
  {volume} {96}},\ \bibinfo {pages} {235139} (\bibinfo {year}
  {2017})}\BibitemShut {NoStop}%
\bibitem [{\citenamefont {Lee}\ \emph {et~al.}(2019)\citenamefont {Lee},
  \citenamefont {Ayral}, \citenamefont {Yao}, \citenamefont {Lanat\`a},\ and\
  \citenamefont {Kotliar}}]{dmet-risb-2}%
  \BibitemOpen
  \bibfield  {author} {\bibinfo {author} {\bibfnamefont {Tsung-Han}\
  \bibnamefont {Lee}}, \bibinfo {author} {\bibfnamefont {Thomas}\ \bibnamefont
  {Ayral}}, \bibinfo {author} {\bibfnamefont {Yong-Xin}\ \bibnamefont {Yao}},
  \bibinfo {author} {\bibfnamefont {Nicola}\ \bibnamefont {Lanat\`a}}, \ and\
  \bibinfo {author} {\bibfnamefont {Gabriel}\ \bibnamefont {Kotliar}},\
  }\bibfield  {title} {\enquote {\bibinfo {title} {Rotationally invariant
  slave-boson and density matrix embedding theory: Unified framework and
  comparative study on the one-dimensional and two-dimensional hubbard
  model},}\ }\href {\doibase 10.1103/PhysRevB.99.115129} {\bibfield  {journal}
  {\bibinfo  {journal} {Phys. Rev. B}\ }\textbf {\bibinfo {volume} {99}},\
  \bibinfo {pages} {115129} (\bibinfo {year} {2019})}\BibitemShut {NoStop}%
\bibitem [{\citenamefont {LeBlanc}\ \emph {et~al.}(2015)\citenamefont
  {LeBlanc}, \citenamefont {Antipov}, \citenamefont {Becca}, \citenamefont
  {Bulik}, \citenamefont {Chan}, \citenamefont {Chung}, \citenamefont {Deng},
  \citenamefont {Ferrero}, \citenamefont {Henderson}, \citenamefont
  {Jim\'enez-Hoyos}, \citenamefont {Kozik}, \citenamefont {Liu}, \citenamefont
  {Millis}, \citenamefont {Prokof'ev}, \citenamefont {Qin}, \citenamefont
  {Scuseria}, \citenamefont {Shi}, \citenamefont {Svistunov}, \citenamefont
  {Tocchio}, \citenamefont {Tupitsyn}, \citenamefont {White}, \citenamefont
  {Zhang}, \citenamefont {Zheng}, \citenamefont {Zhu},\ and\ \citenamefont
  {Gull}}]{many-qmc}%
  \BibitemOpen
  \bibfield  {author} {\bibinfo {author} {\bibfnamefont {J.~P.~F.}\
  \bibnamefont {LeBlanc}}, \bibinfo {author} {\bibfnamefont {Andrey~E.}\
  \bibnamefont {Antipov}}, \bibinfo {author} {\bibfnamefont {Federico}\
  \bibnamefont {Becca}}, \bibinfo {author} {\bibfnamefont {Ireneusz~W.}\
  \bibnamefont {Bulik}}, \bibinfo {author} {\bibfnamefont {Garnet Kin-Lic}\
  \bibnamefont {Chan}}, \bibinfo {author} {\bibfnamefont {Chia-Min}\
  \bibnamefont {Chung}}, \bibinfo {author} {\bibfnamefont {Youjin}\
  \bibnamefont {Deng}}, \bibinfo {author} {\bibfnamefont {Michel}\ \bibnamefont
  {Ferrero}}, \bibinfo {author} {\bibfnamefont {Thomas~M.}\ \bibnamefont
  {Henderson}}, \bibinfo {author} {\bibfnamefont {Carlos~A.}\ \bibnamefont
  {Jim\'enez-Hoyos}}, \bibinfo {author} {\bibfnamefont {E.}~\bibnamefont
  {Kozik}}, \bibinfo {author} {\bibfnamefont {Xuan-Wen}\ \bibnamefont {Liu}},
  \bibinfo {author} {\bibfnamefont {Andrew~J.}\ \bibnamefont {Millis}},
  \bibinfo {author} {\bibfnamefont {N.~V.}\ \bibnamefont {Prokof'ev}}, \bibinfo
  {author} {\bibfnamefont {Mingpu}\ \bibnamefont {Qin}}, \bibinfo {author}
  {\bibfnamefont {Gustavo~E.}\ \bibnamefont {Scuseria}}, \bibinfo {author}
  {\bibfnamefont {Hao}\ \bibnamefont {Shi}}, \bibinfo {author} {\bibfnamefont
  {B.~V.}\ \bibnamefont {Svistunov}}, \bibinfo {author} {\bibfnamefont
  {Luca~F.}\ \bibnamefont {Tocchio}}, \bibinfo {author} {\bibfnamefont {I.~S.}\
  \bibnamefont {Tupitsyn}}, \bibinfo {author} {\bibfnamefont {Steven~R.}\
  \bibnamefont {White}}, \bibinfo {author} {\bibfnamefont {Shiwei}\
  \bibnamefont {Zhang}}, \bibinfo {author} {\bibfnamefont {Bo-Xiao}\
  \bibnamefont {Zheng}}, \bibinfo {author} {\bibfnamefont {Zhenyue}\
  \bibnamefont {Zhu}}, \ and\ \bibinfo {author} {\bibfnamefont {Emanuel}\
  \bibnamefont {Gull}} (\bibinfo {collaboration} {Simons Collaboration on the
  Many-Electron Problem}),\ }\bibfield  {title} {\enquote {\bibinfo {title}
  {Solutions of the two-dimensional hubbard model: Benchmarks and results from
  a wide range of numerical algorithms},}\ }\href@noop {} {\bibfield  {journal}
  {\bibinfo  {journal} {Phys. Rev. X}\ }\textbf {\bibinfo {volume} {5}},\
  \bibinfo {pages} {041041} (\bibinfo {year} {2015})}\BibitemShut {NoStop}%
\bibitem [{\citenamefont {Wilson}(1975)}]{Wilson}%
  \BibitemOpen
  \bibfield  {author} {\bibinfo {author} {\bibfnamefont {Kenneth~G.}\
  \bibnamefont {Wilson}},\ }\href@noop {} {\bibfield  {journal} {\bibinfo
  {journal} {Rev. Mod. Phys.}\ }\textbf {\bibinfo {volume} {47}},\ \bibinfo
  {pages} {773} (\bibinfo {year} {1975})}\BibitemShut {NoStop}%
\bibitem [{\citenamefont {White}(1992)}]{DMRG-original-White-PRL}%
  \BibitemOpen
  \bibfield  {author} {\bibinfo {author} {\bibfnamefont {Steven~R.}\
  \bibnamefont {White}},\ }\bibfield  {title} {\enquote {\bibinfo {title}
  {Density matrix formulation for quantum renormalization groups},}\
  }\href@noop {} {\bibfield  {journal} {\bibinfo  {journal} {Phys. Rev. Lett.}\
  }\textbf {\bibinfo {volume} {69}},\ \bibinfo {pages} {2863--2866} (\bibinfo
  {year} {1992})}\BibitemShut {NoStop}%
\bibitem [{\citenamefont {Weichselbaum}\ \emph {et~al.}(2009)\citenamefont
  {Weichselbaum}, \citenamefont {Verstraete}, \citenamefont {Schollw\"ock},
  \citenamefont {Cirac},\ and\ \citenamefont {von
  Delft}}]{Vaestrate_NRG-DMRG-MPS}%
  \BibitemOpen
  \bibfield  {author} {\bibinfo {author} {\bibfnamefont {A.}~\bibnamefont
  {Weichselbaum}}, \bibinfo {author} {\bibfnamefont {F.}~\bibnamefont
  {Verstraete}}, \bibinfo {author} {\bibfnamefont {U.}~\bibnamefont
  {Schollw\"ock}}, \bibinfo {author} {\bibfnamefont {J.~I.}\ \bibnamefont
  {Cirac}}, \ and\ \bibinfo {author} {\bibfnamefont {Jan}\ \bibnamefont {von
  Delft}},\ }\bibfield  {title} {\enquote {\bibinfo {title} {Variational
  matrix-product-state approach to quantum impurity models},}\ }\href@noop {}
  {\bibfield  {journal} {\bibinfo  {journal} {Phys. Rev. B}\ }\textbf {\bibinfo
  {volume} {80}},\ \bibinfo {pages} {165117} (\bibinfo {year}
  {2009})}\BibitemShut {NoStop}%
\bibitem [{\citenamefont {Stuart}\ \emph {et~al.}(1997)\citenamefont {Stuart},
  \citenamefont {Susan~J.},\ and\ \citenamefont {Joel~M.}}]{n-mode-1}%
  \BibitemOpen
  \bibfield  {author} {\bibinfo {author} {\bibfnamefont {Carter}\ \bibnamefont
  {Stuart}}, \bibinfo {author} {\bibfnamefont {Culik}\ \bibnamefont
  {Susan~J.}}, \ and\ \bibinfo {author} {\bibfnamefont {Bowman}\ \bibnamefont
  {Joel~M.}},\ }\bibfield  {title} {\enquote {\bibinfo {title} {Vibrational
  self-consistent field method for many-mode systems: A new approach and
  application to the vibrations of $\text{CO}$ adsorbed on $\text{Cu(100)}$},}\
  }\href@noop {} {\bibfield  {journal} {\bibinfo  {journal} {J. Chem. Phys.}\
  }\textbf {\bibinfo {volume} {107}},\ \bibinfo {pages} {10458} (\bibinfo
  {year} {1997})}\BibitemShut {NoStop}%
\bibitem [{\citenamefont {Kun}\ \emph {et~al.}(2017)\citenamefont {Kun},
  \citenamefont {John~E.},\ and\ \citenamefont {John}}]{n-mode-2}%
  \BibitemOpen
  \bibfield  {author} {\bibinfo {author} {\bibfnamefont {Yao}\ \bibnamefont
  {Kun}}, \bibinfo {author} {\bibfnamefont {Herr}\ \bibnamefont {John~E.}}, \
  and\ \bibinfo {author} {\bibfnamefont {Parkhill}\ \bibnamefont {John}},\
  }\bibfield  {title} {\enquote {\bibinfo {title} {The many-body expansion
  combined with neural networks},}\ }\href@noop {} {\bibfield  {journal}
  {\bibinfo  {journal} {J. Chem. Phys.}\ }\textbf {\bibinfo {volume} {146}},\
  \bibinfo {pages} {014106} (\bibinfo {year} {2017})}\BibitemShut {NoStop}%
\bibitem [{\citenamefont {Emil~Lund}\ \emph {et~al.}(2018)\citenamefont
  {Emil~Lund}, \citenamefont {Bo}, \citenamefont {Ian~Heide},\ and\
  \citenamefont {Ove}}]{n-mode-3}%
  \BibitemOpen
  \bibfield  {author} {\bibinfo {author} {\bibfnamefont {Klinting}\
  \bibnamefont {Emil~Lund}}, \bibinfo {author} {\bibfnamefont {Thomsen}\
  \bibnamefont {Bo}}, \bibinfo {author} {\bibfnamefont {Godtliebsen}\
  \bibnamefont {Ian~Heide}}, \ and\ \bibinfo {author} {\bibfnamefont
  {Christiansen}\ \bibnamefont {Ove}},\ }\bibfield  {title} {\enquote {\bibinfo
  {title} {Employing general fit-bases for construction of potential energy
  surfaces with an adaptive density-guided approach},}\ }\href@noop {}
  {\bibfield  {journal} {\bibinfo  {journal} {J. Chem. Phys.}\ }\textbf
  {\bibinfo {volume} {148}},\ \bibinfo {pages} {064113} (\bibinfo {year}
  {2018})}\BibitemShut {NoStop}%
\bibitem [{\citenamefont {Carolin}\ and\ \citenamefont {Ove}(2016)}]{n-mode-4}%
  \BibitemOpen
  \bibfield  {author} {\bibinfo {author} {\bibfnamefont {K\"onig}\ \bibnamefont
  {Carolin}}\ and\ \bibinfo {author} {\bibfnamefont {Christiansen}\
  \bibnamefont {Ove}},\ }\bibfield  {title} {\enquote {\bibinfo {title}
  {Linear-scaling generation of potential energy surfaces using a double
  incremental expansion},}\ }\href@noop {} {\bibfield  {journal} {\bibinfo
  {journal} {J. Chem. Phys.}\ }\textbf {\bibinfo {volume} {145}},\ \bibinfo
  {pages} {064105} (\bibinfo {year} {2016})}\BibitemShut {NoStop}%
\bibitem [{\citenamefont {Schmidt}\ \emph {et~al.}(2019)\citenamefont
  {Schmidt}, \citenamefont {Marques}, \citenamefont {Botti},\ and\
  \citenamefont {Marques}}]{ML-Schmidt2019}%
  \BibitemOpen
  \bibfield  {author} {\bibinfo {author} {\bibfnamefont {Jonathan}\
  \bibnamefont {Schmidt}}, \bibinfo {author} {\bibfnamefont {M{\'a}rio R.~G.}\
  \bibnamefont {Marques}}, \bibinfo {author} {\bibfnamefont {Silvana}\
  \bibnamefont {Botti}}, \ and\ \bibinfo {author} {\bibfnamefont {Miguel
  A.~L.}\ \bibnamefont {Marques}},\ }\bibfield  {title} {\enquote {\bibinfo
  {title} {Recent advances and applications of machine learning in solid-state
  materials science},}\ }\href {\doibase 10.1038/s41524-019-0221-0} {\bibfield
  {journal} {\bibinfo  {journal} {npj Computational Materials}\ }\textbf
  {\bibinfo {volume} {5}},\ \bibinfo {pages} {83} (\bibinfo {year}
  {2019})}\BibitemShut {NoStop}%
\bibitem [{\citenamefont {Albers}(2001)}]{Pu-nature-albers-2001}%
  \BibitemOpen
  \bibfield  {author} {\bibinfo {author} {\bibfnamefont {R.~C.}\ \bibnamefont
  {Albers}},\ }\bibfield  {title} {\enquote {\bibinfo {title} {An expanding
  view of plutonium},}\ }\href@noop {} {\bibfield  {journal} {\bibinfo
  {journal} {Nature}\ }\textbf {\bibinfo {volume} {410}},\ \bibinfo {pages}
  {759} (\bibinfo {year} {2001})}\BibitemShut {NoStop}%
\bibitem [{\citenamefont {Amadon}(2016)}]{Amadon-actinides}%
  \BibitemOpen
  \bibfield  {author} {\bibinfo {author} {\bibfnamefont {Bernard}\ \bibnamefont
  {Amadon}},\ }\bibfield  {title} {\enquote {\bibinfo {title} {First-principles
  $\text{DFT+DMFT}$ calculations of structural properties of actinides: Role of
  $\text{Hund}$'s exchange, spin-orbit coupling, and crystal structure},}\
  }\href@noop {} {\bibfield  {journal} {\bibinfo  {journal} {Phys. Rev. B}\
  }\textbf {\bibinfo {volume} {94}},\ \bibinfo {pages} {115148} (\bibinfo
  {year} {2016})}\BibitemShut {NoStop}%
\bibitem [{\citenamefont {Zhu}\ \emph {et~al.}(2013)\citenamefont {Zhu},
  \citenamefont {Albers}, \citenamefont {Haule}, \citenamefont {Kotliar},\ and\
  \citenamefont {Wills}}]{alphaPu-Gabi-nature-2013}%
  \BibitemOpen
  \bibfield  {author} {\bibinfo {author} {\bibfnamefont {J.~X.}\ \bibnamefont
  {Zhu}}, \bibinfo {author} {\bibfnamefont {R.~C.}\ \bibnamefont {Albers}},
  \bibinfo {author} {\bibfnamefont {K.}~\bibnamefont {Haule}}, \bibinfo
  {author} {\bibfnamefont {G.}~\bibnamefont {Kotliar}}, \ and\ \bibinfo
  {author} {\bibfnamefont {J.~M.}\ \bibnamefont {Wills}},\ }\bibfield  {title}
  {\enquote {\bibinfo {title} {Site-selective electronic correlation in
  $\alpha$-plutonium metal},}\ }\href@noop {} {\bibfield  {journal} {\bibinfo
  {journal} {Nat. Commun.}\ }\textbf {\bibinfo {volume} {4:2644}} (\bibinfo
  {year} {2013})}\BibitemShut {NoStop}%
\bibitem [{\citenamefont {Savrasov}\ \emph {et~al.}(2001)\citenamefont
  {Savrasov}, \citenamefont {Kotliar},\ and\ \citenamefont
  {Abrahams}}]{Pu-Gabi-nature-2001}%
  \BibitemOpen
  \bibfield  {author} {\bibinfo {author} {\bibfnamefont {S.~Y.}\ \bibnamefont
  {Savrasov}}, \bibinfo {author} {\bibfnamefont {G.}~\bibnamefont {Kotliar}}, \
  and\ \bibinfo {author} {\bibfnamefont {E.}~\bibnamefont {Abrahams}},\
  }\bibfield  {title} {\enquote {\bibinfo {title} {Correlated electrons in
  $\delta$-plutonium within a dynamical mean-field picture},}\ }\href@noop {}
  {\bibfield  {journal} {\bibinfo  {journal} {Nature}\ }\textbf {\bibinfo
  {volume} {410}},\ \bibinfo {pages} {793} (\bibinfo {year}
  {2001})}\BibitemShut {NoStop}%
\bibitem [{\citenamefont {Pourovskii}\ \emph {et~al.}(2007)\citenamefont
  {Pourovskii}, \citenamefont {Kotliar}, \citenamefont {Katsnelson},\ and\
  \citenamefont {Lichtenstein}}]{sh-alpha-delta-Pu}%
  \BibitemOpen
  \bibfield  {author} {\bibinfo {author} {\bibfnamefont {L.~V.}\ \bibnamefont
  {Pourovskii}}, \bibinfo {author} {\bibfnamefont {G.}~\bibnamefont {Kotliar}},
  \bibinfo {author} {\bibfnamefont {M.~I.}\ \bibnamefont {Katsnelson}}, \ and\
  \bibinfo {author} {\bibfnamefont {A.~I.}\ \bibnamefont {Lichtenstein}},\
  }\bibfield  {title} {\enquote {\bibinfo {title} {Dynamical mean-field theory
  investigation of specific heat and electronic structure of $\alpha$- and
  $\delta$-plutonium},}\ }\href@noop {} {\bibfield  {journal} {\bibinfo
  {journal} {Phys. Rev. B}\ }\textbf {\bibinfo {volume} {75}},\ \bibinfo
  {pages} {235107} (\bibinfo {year} {2007})}\BibitemShut {NoStop}%
\bibitem [{\citenamefont {Anisimov}\ \emph
  {et~al.}(1997{\natexlab{b}})\citenamefont {Anisimov}, \citenamefont
  {Aryasetiawan},\ and\ \citenamefont {Lichtenstein}}]{LDA+U}%
  \BibitemOpen
  \bibfield  {author} {\bibinfo {author} {\bibfnamefont {V.~I.}\ \bibnamefont
  {Anisimov}}, \bibinfo {author} {\bibfnamefont {F.}~\bibnamefont
  {Aryasetiawan}}, \ and\ \bibinfo {author} {\bibfnamefont {A.~I.}\
  \bibnamefont {Lichtenstein}},\ }\bibfield  {title} {\enquote {\bibinfo
  {title} {First-principles calculations of the electronic structure and
  spectra of strongly correlated systems: the $\text{LDA+U}$ method},}\
  }\href@noop {} {\bibfield  {journal} {\bibinfo  {journal} {J. Phys. Condens.
  Matter}\ }\textbf {\bibinfo {volume} {9}},\ \bibinfo {pages} {767} (\bibinfo
  {year} {1997}{\natexlab{b}})}\BibitemShut {NoStop}%
\bibitem [{\citenamefont {Kanungo}\ \emph {et~al.}(2019)\citenamefont
  {Kanungo}, \citenamefont {Zimmerman},\ and\ \citenamefont
  {Gavini}}]{Kanungo2019}%
  \BibitemOpen
  \bibfield  {author} {\bibinfo {author} {\bibfnamefont {Bikash}\ \bibnamefont
  {Kanungo}}, \bibinfo {author} {\bibfnamefont {Paul~M.}\ \bibnamefont
  {Zimmerman}}, \ and\ \bibinfo {author} {\bibfnamefont {Vikram}\ \bibnamefont
  {Gavini}},\ }\bibfield  {title} {\enquote {\bibinfo {title} {Exact
  exchange-correlation potentials from ground-state electron densities},}\
  }\href@noop {} {\bibfield  {journal} {\bibinfo  {journal} {Nat. Commun.}\
  }\textbf {\bibinfo {volume} {10}},\ \bibinfo {pages} {4497} (\bibinfo {year}
  {2019})}\BibitemShut {NoStop}%
\bibitem [{\citenamefont {Bart\'ok}\ \emph {et~al.}(2010)\citenamefont
  {Bart\'ok}, \citenamefont {Payne}, \citenamefont {Kondor},\ and\
  \citenamefont {Cs\'anyi}}]{surrogate-1}%
  \BibitemOpen
  \bibfield  {author} {\bibinfo {author} {\bibfnamefont {Albert~P.}\
  \bibnamefont {Bart\'ok}}, \bibinfo {author} {\bibfnamefont {Mike~C.}\
  \bibnamefont {Payne}}, \bibinfo {author} {\bibfnamefont {Risi}\ \bibnamefont
  {Kondor}}, \ and\ \bibinfo {author} {\bibfnamefont {G\'abor}\ \bibnamefont
  {Cs\'anyi}},\ }\bibfield  {title} {\enquote {\bibinfo {title} {Gaussian
  approximation potentials: The accuracy of quantum mechanics, without the
  electrons},}\ }\href@noop {} {\bibfield  {journal} {\bibinfo  {journal}
  {Phys. Rev. Lett.}\ }\textbf {\bibinfo {volume} {104}},\ \bibinfo {pages}
  {136403} (\bibinfo {year} {2010})}\BibitemShut {NoStop}%
\bibitem [{\citenamefont {Chandramouli}\ \emph {et~al.}(2019)\citenamefont
  {Chandramouli}, \citenamefont {Matthias}, \citenamefont {Brayden},
  \citenamefont {Alexander~V.}, \citenamefont {Tim}, \citenamefont {Conrad~W.},
  \citenamefont {Gábor}, \citenamefont {David~W.},\ and\ \citenamefont
  {Gus~L.~W.}}]{surrogate-2}%
  \BibitemOpen
  \bibfield  {author} {\bibinfo {author} {\bibfnamefont {Nyshadham}\
  \bibnamefont {Chandramouli}}, \bibinfo {author} {\bibfnamefont {Rupp}\
  \bibnamefont {Matthias}}, \bibinfo {author} {\bibfnamefont {Bekker}\
  \bibnamefont {Brayden}}, \bibinfo {author} {\bibfnamefont {Shapeev}\
  \bibnamefont {Alexander~V.}}, \bibinfo {author} {\bibfnamefont {Mueller}\
  \bibnamefont {Tim}}, \bibinfo {author} {\bibfnamefont {Rosenbrock}\
  \bibnamefont {Conrad~W.}}, \bibinfo {author} {\bibfnamefont {Csányi}\
  \bibnamefont {Gábor}}, \bibinfo {author} {\bibfnamefont {Wingate}\
  \bibnamefont {David~W.}}, \ and\ \bibinfo {author} {\bibfnamefont {Hart}\
  \bibnamefont {Gus~L.~W.}},\ }\bibfield  {title} {\enquote {\bibinfo {title}
  {Machine-learned multi-system surrogate models for materials prediction},}\
  }\href@noop {} {\bibfield  {journal} {\bibinfo  {journal} {npj Comput.
  Mater.}\ }\textbf {\bibinfo {volume} {5}},\ \bibinfo {pages} {51} (\bibinfo
  {year} {2019})}\BibitemShut {NoStop}%
\bibitem [{\citenamefont {Felix}\ \emph {et~al.}(2017)\citenamefont {Felix},
  \citenamefont {Leslie}, \citenamefont {Li}, \citenamefont {Mark~E.},
  \citenamefont {Kieron},\ and\ \citenamefont
  {Klaus-Robert}}]{surrogate-DFT-1}%
  \BibitemOpen
  \bibfield  {author} {\bibinfo {author} {\bibfnamefont {Brockherde}\
  \bibnamefont {Felix}}, \bibinfo {author} {\bibfnamefont {Vogt}\ \bibnamefont
  {Leslie}}, \bibinfo {author} {\bibfnamefont {Li}~\bibnamefont {Li}}, \bibinfo
  {author} {\bibfnamefont {Tuckerman}\ \bibnamefont {Mark~E.}}, \bibinfo
  {author} {\bibfnamefont {Burke}\ \bibnamefont {Kieron}}, \ and\ \bibinfo
  {author} {\bibfnamefont {M\"uller}\ \bibnamefont {Klaus-Robert}},\ }\bibfield
   {title} {\enquote {\bibinfo {title} {Bypassing the kohn-sham equations with
  machine learning},}\ }\href@noop {} {\bibfield  {journal} {\bibinfo
  {journal} {Nat. Commun.}\ }\textbf {\bibinfo {volume} {8}},\ \bibinfo {pages}
  {872} (\bibinfo {year} {2017})}\BibitemShut {NoStop}%
\bibitem [{\citenamefont {Li}\ \emph {et~al.}(2016)\citenamefont {Li},
  \citenamefont {Baker}, \citenamefont {White},\ and\ \citenamefont
  {Burke}}]{surrogate-DFT-2}%
  \BibitemOpen
  \bibfield  {author} {\bibinfo {author} {\bibfnamefont {Li}~\bibnamefont
  {Li}}, \bibinfo {author} {\bibfnamefont {Thomas~E.}\ \bibnamefont {Baker}},
  \bibinfo {author} {\bibfnamefont {Steven~R.}\ \bibnamefont {White}}, \ and\
  \bibinfo {author} {\bibfnamefont {Kieron}\ \bibnamefont {Burke}},\ }\bibfield
   {title} {\enquote {\bibinfo {title} {Pure density functional for strong
  correlation and the thermodynamic limit from machine learning},}\ }\href@noop
  {} {\bibfield  {journal} {\bibinfo  {journal} {Phys. Rev. B}\ }\textbf
  {\bibinfo {volume} {94}},\ \bibinfo {pages} {245129} (\bibinfo {year}
  {2016})}\BibitemShut {NoStop}%
\bibitem [{\citenamefont {Arsenault}\ \emph {et~al.}(2014)\citenamefont
  {Arsenault}, \citenamefont {Lopez-Bezanilla}, \citenamefont {von
  Lilienfeld},\ and\ \citenamefont {Millis}}]{surrogate-Millis}%
  \BibitemOpen
  \bibfield  {author} {\bibinfo {author} {\bibfnamefont {Louis-Fran\ifmmode
  \mbox{\c{c}}\else~\c{c}\fi{}ois}\ \bibnamefont {Arsenault}}, \bibinfo
  {author} {\bibfnamefont {Alejandro}\ \bibnamefont {Lopez-Bezanilla}},
  \bibinfo {author} {\bibfnamefont {O.~Anatole}\ \bibnamefont {von
  Lilienfeld}}, \ and\ \bibinfo {author} {\bibfnamefont {Andrew~J.}\
  \bibnamefont {Millis}},\ }\bibfield  {title} {\enquote {\bibinfo {title}
  {Machine learning for many-body physics: The case of the anderson impurity
  model},}\ }\href {\doibase 10.1103/PhysRevB.90.155136} {\bibfield  {journal}
  {\bibinfo  {journal} {Phys. Rev. B}\ }\textbf {\bibinfo {volume} {90}},\
  \bibinfo {pages} {155136} (\bibinfo {year} {2014})}\BibitemShut {NoStop}%
\bibitem [{\citenamefont {Keith~T.}\ \emph {et~al.}(2018)\citenamefont
  {Keith~T.}, \citenamefont {Daniel~W.}, \citenamefont {Hugh}, \citenamefont
  {Olexandr},\ and\ \citenamefont {Aron}}]{ML-materials-1}%
  \BibitemOpen
  \bibfield  {author} {\bibinfo {author} {\bibfnamefont {Butler}\ \bibnamefont
  {Keith~T.}}, \bibinfo {author} {\bibfnamefont {Davies}\ \bibnamefont
  {Daniel~W.}}, \bibinfo {author} {\bibfnamefont {Cartwright}\ \bibnamefont
  {Hugh}}, \bibinfo {author} {\bibfnamefont {Isayev}\ \bibnamefont {Olexandr}},
  \ and\ \bibinfo {author} {\bibfnamefont {Walsh}\ \bibnamefont {Aron}},\
  }\bibfield  {title} {\enquote {\bibinfo {title} {Machine learning for
  molecular and materials science},}\ }\href@noop {} {\bibfield  {journal}
  {\bibinfo  {journal} {Nature}\ }\textbf {\bibinfo {volume} {559}},\ \bibinfo
  {pages} {547} (\bibinfo {year} {2018})}\BibitemShut {NoStop}%
\bibitem [{\citenamefont {Jung}\ and\ \citenamefont
  {Gerber}(1996)}]{jung_vibrational_1996}%
  \BibitemOpen
  \bibfield  {author} {\bibinfo {author} {\bibfnamefont {Joon~O.}\ \bibnamefont
  {Jung}}\ and\ \bibinfo {author} {\bibfnamefont {R.~Benny}\ \bibnamefont
  {Gerber}},\ }\bibfield  {title} {\enquote {\bibinfo {title} {Vibrational wave
  functions and spectroscopy of ($\text{H}_2\text{O}$)$_n$, $n=2,3,4,5$:
  {Vibrational} self-consistent field with correlation corrections},}\ }\href
  {\doibase 10.1063/1.472960} {\bibfield  {journal} {\bibinfo  {journal} {The
  Journal of Chemical Physics}\ }\textbf {\bibinfo {volume} {105}},\ \bibinfo
  {pages} {10332--10348} (\bibinfo {year} {1996})},\ \bibinfo {note}
  {publisher: American Institute of Physics}\BibitemShut {NoStop}%
\bibitem [{\citenamefont {Zimmerman}(2017)}]{zimmerman_strong_2017}%
  \BibitemOpen
  \bibfield  {author} {\bibinfo {author} {\bibfnamefont {Paul~M.}\ \bibnamefont
  {Zimmerman}},\ }\bibfield  {title} {\enquote {\bibinfo {title} {Strong
  correlation in incremental full configuration interaction},}\ }\href
  {\doibase 10.1063/1.4985566} {\bibfield  {journal} {\bibinfo  {journal} {The
  Journal of Chemical Physics}\ }\textbf {\bibinfo {volume} {146}},\ \bibinfo
  {pages} {224104} (\bibinfo {year} {2017})},\ \bibinfo {note} {publisher:
  American Institute of Physics}\BibitemShut {NoStop}%
\bibitem [{\citenamefont {Stoll}(2019)}]{stoll_toward_2019}%
  \BibitemOpen
  \bibfield  {author} {\bibinfo {author} {\bibfnamefont {Hermann}\ \bibnamefont
  {Stoll}},\ }\bibfield  {title} {\enquote {\bibinfo {title} {Toward a
  wavefunction-based treatment of strong electron correlation in extended
  systems by means of incremental methods},}\ }\href {\doibase
  10.1063/1.5109860} {\bibfield  {journal} {\bibinfo  {journal} {The Journal of
  Chemical Physics}\ }\textbf {\bibinfo {volume} {151}},\ \bibinfo {pages}
  {044104} (\bibinfo {year} {2019})},\ \bibinfo {note} {publisher: American
  Institute of Physics}\BibitemShut {NoStop}%
\bibitem [{\citenamefont {Lanat\`a}\ \emph {et~al.}(2012)\citenamefont
  {Lanat\`a}, \citenamefont {Strand}, \citenamefont {Dai},\ and\ \citenamefont
  {Hellsing}}]{Gmethod}%
  \BibitemOpen
  \bibfield  {author} {\bibinfo {author} {\bibfnamefont {N.}~\bibnamefont
  {Lanat\`a}}, \bibinfo {author} {\bibfnamefont {H.~U.~R.}\ \bibnamefont
  {Strand}}, \bibinfo {author} {\bibfnamefont {X.}~\bibnamefont {Dai}}, \ and\
  \bibinfo {author} {\bibfnamefont {B.}~\bibnamefont {Hellsing}},\ }\bibfield
  {title} {\enquote {\bibinfo {title} {Efficient implementation of the
  $\text{Gutzwiller}$ variational method},}\ }\href {\doibase
  10.1103/PhysRevB.85.035133} {\bibfield  {journal} {\bibinfo  {journal} {Phys.
  Rev. B}\ }\textbf {\bibinfo {volume} {85}},\ \bibinfo {pages} {035133}
  (\bibinfo {year} {2012})}\BibitemShut {NoStop}%
\bibitem [{\citenamefont {Moore}\ and\ \citenamefont {van~der
  Laan}(2009)}]{Actinide_metals-review}%
  \BibitemOpen
  \bibfield  {author} {\bibinfo {author} {\bibfnamefont {Kevin~T.}\
  \bibnamefont {Moore}}\ and\ \bibinfo {author} {\bibfnamefont {Gerrit}\
  \bibnamefont {van~der Laan}},\ }\bibfield  {title} {\enquote {\bibinfo
  {title} {Nature of the $5f$ states in actinide metals},}\ }\href {\doibase
  10.1103/RevModPhys.81.235} {\bibfield  {journal} {\bibinfo  {journal} {Rev.
  Mod. Phys.}\ }\textbf {\bibinfo {volume} {81}},\ \bibinfo {pages} {235--298}
  (\bibinfo {year} {2009})}\BibitemShut {NoStop}%
\bibitem [{\citenamefont {Zachariasen}(1959)}]{Zachariasen-Pa}%
  \BibitemOpen
  \bibfield  {author} {\bibinfo {author} {\bibfnamefont {W.~H.}\ \bibnamefont
  {Zachariasen}},\ }\bibfield  {title} {\enquote {\bibinfo {title} {{On the
  crystal structure of protactinium metal}},}\ }\href@noop {} {\bibfield
  {journal} {\bibinfo  {journal} {Acta Crystallographica}\ }\textbf {\bibinfo
  {volume} {12}},\ \bibinfo {pages} {698--700} (\bibinfo {year}
  {1959})}\BibitemShut {NoStop}%
\bibitem [{\citenamefont {Schwarz}\ and\ \citenamefont {Blaha}(2003)}]{WIEN2k}%
  \BibitemOpen
  \bibfield  {author} {\bibinfo {author} {\bibfnamefont {Karlheinz}\
  \bibnamefont {Schwarz}}\ and\ \bibinfo {author} {\bibfnamefont {Peter}\
  \bibnamefont {Blaha}},\ }\bibfield  {title} {\enquote {\bibinfo {title}
  {Solid state calculations using {WIEN2k}},}\ }\href@noop {} {\bibfield
  {journal} {\bibinfo  {journal} {Computational Materials Science}\ }\textbf
  {\bibinfo {volume} {28}},\ \bibinfo {pages} {259 -- 273} (\bibinfo {year}
  {2003})}\BibitemShut {NoStop}%
\end{thebibliography}

\begin{thebibliography}{17}%
\makeatletter
\providecommand \@ifxundefined [1]{%
 \@ifx{#1\undefined}
}%
\providecommand \@ifnum [1]{%
 \ifnum #1\expandafter \@firstoftwo
 \else \expandafter \@secondoftwo
 \fi
}%
\providecommand \@ifx [1]{%
 \ifx #1\expandafter \@firstoftwo
 \else \expandafter \@secondoftwo
 \fi
}%
\providecommand \natexlab [1]{#1}%
\providecommand \enquote  [1]{``#1''}%
\providecommand \bibnamefont  [1]{#1}%
\providecommand \bibfnamefont [1]{#1}%
\providecommand \citenamefont [1]{#1}%
\providecommand \href@noop [0]{\@secondoftwo}%
\providecommand \href [0]{\begingroup \@sanitize@url \@href}%
\providecommand \@href[1]{\@@startlink{#1}\@@href}%
\providecommand \@@href[1]{\endgroup#1\@@endlink}%
\providecommand \@sanitize@url [0]{\catcode `\\12\catcode `\$12\catcode
  `\&12\catcode `\#12\catcode `\^12\catcode `\_12\catcode `\%12\relax}%
\providecommand \@@startlink[1]{}%
\providecommand \@@endlink[0]{}%
\providecommand \url  [0]{\begingroup\@sanitize@url \@url }%
\providecommand \@url [1]{\endgroup\@href {#1}{\urlprefix }}%
\providecommand \urlprefix  [0]{URL }%
\providecommand \Eprint [0]{\href }%
\providecommand \doibase [0]{http://dx.doi.org/}%
\providecommand \selectlanguage [0]{\@gobble}%
\providecommand \bibinfo  [0]{\@secondoftwo}%
\providecommand \bibfield  [0]{\@secondoftwo}%
\providecommand \translation [1]{[#1]}%
\providecommand \BibitemOpen [0]{}%
\providecommand \bibitemStop [0]{}%
\providecommand \bibitemNoStop [0]{.\EOS\space}%
\providecommand \EOS [0]{\spacefactor3000\relax}%
\providecommand \BibitemShut  [1]{\csname bibitem#1\endcsname}%
\let\auto@bib@innerbib\@empty
%</preamble>
\bibitem [{\citenamefont {Lanat\`a}\ \emph {et~al.}(2015)\citenamefont
  {Lanat\`a}, \citenamefont {Yao}, \citenamefont {Wang}, \citenamefont {Ho},\
  and\ \citenamefont {Kotliar}}]{Our-PRXsupp}%
  \BibitemOpen
  \bibfield  {author} {\bibinfo {author} {\bibfnamefont {Nicola}\ \bibnamefont
  {Lanat\`a}}, \bibinfo {author} {\bibfnamefont {Yong~Xin}\ \bibnamefont
  {Yao}}, \bibinfo {author} {\bibfnamefont {Cai-Zhuang}\ \bibnamefont {Wang}},
  \bibinfo {author} {\bibfnamefont {Kai-Ming}\ \bibnamefont {Ho}}, \ and\
  \bibinfo {author} {\bibfnamefont {Gabriel}\ \bibnamefont {Kotliar}},\
  }\bibfield  {title} {\enquote {\bibinfo {title} {Phase diagram and electronic
  structure of praseodymium and plutonium},}\ }\href@noop {} {\bibfield
  {journal} {\bibinfo  {journal} {Phys. Rev. X}\ }\textbf {\bibinfo {volume}
  {5}},\ \bibinfo {pages} {011008} (\bibinfo {year} {2015})}\BibitemShut
  {NoStop}%
\bibitem [{\citenamefont {Lanat{\`{a}}}\ \emph {et~al.}(2017)\citenamefont
  {Lanat{\`{a}}}, \citenamefont {Yao}, \citenamefont {Deng}, \citenamefont
  {Dobrosavljevi{\'{c}}},\ and\ \citenamefont {Kotliar}}]{Lanata2016supp}%
  \BibitemOpen
  \bibfield  {author} {\bibinfo {author} {\bibfnamefont {Nicola}\ \bibnamefont
  {Lanat{\`{a}}}}, \bibinfo {author} {\bibfnamefont {Yongxin}\ \bibnamefont
  {Yao}}, \bibinfo {author} {\bibfnamefont {Xiaoyu}\ \bibnamefont {Deng}},
  \bibinfo {author} {\bibfnamefont {Vladimir}\ \bibnamefont
  {Dobrosavljevi{\'{c}}}}, \ and\ \bibinfo {author} {\bibfnamefont {Gabriel}\
  \bibnamefont {Kotliar}},\ }\bibfield  {title} {\enquote {\bibinfo {title}
  {{Slave Boson Theory of Orbital Differentiation with Crystal Field Effects:
  Application to UO$_2$}},}\ }\href {\doibase 10.1103/PhysRevLett.118.126401}
  {\bibfield  {journal} {\bibinfo  {journal} {Physical Review Letters}\
  }\textbf {\bibinfo {volume} {118}},\ \bibinfo {pages} {126401} (\bibinfo
  {year} {2017})}\BibitemShut {NoStop}%
\bibitem [{\citenamefont {B\"unemann}\ \emph {et~al.}(2003)\citenamefont
  {B\"unemann}, \citenamefont {Gebhard},\ and\ \citenamefont
  {Thul}}]{Gebhard-FLsupp}%
  \BibitemOpen
  \bibfield  {author} {\bibinfo {author} {\bibfnamefont {J.}~\bibnamefont
  {B\"unemann}}, \bibinfo {author} {\bibfnamefont {F.}~\bibnamefont {Gebhard}},
  \ and\ \bibinfo {author} {\bibfnamefont {R.}~\bibnamefont {Thul}},\
  }\bibfield  {title} {\enquote {\bibinfo {title} {Landau-$\text{Gutzwiller}$
  quasiparticles},}\ }\href@noop {} {\bibfield  {journal} {\bibinfo  {journal}
  {Phys. Rev. B}\ }\textbf {\bibinfo {volume} {67}},\ \bibinfo {pages} {075103}
  (\bibinfo {year} {2003})}\BibitemShut {NoStop}%
\bibitem [{\citenamefont {Schwarz}\ and\ \citenamefont {Blaha}(2003)}]{WIEN2ksupp}%
  \BibitemOpen
  \bibfield  {author} {\bibinfo {author} {\bibfnamefont {Karlheinz}\
  \bibnamefont {Schwarz}}\ and\ \bibinfo {author} {\bibfnamefont {Peter}\
  \bibnamefont {Blaha}},\ }\bibfield  {title} {\enquote {\bibinfo {title}
  {Solid state calculations using {WIEN2k}},}\ }\href@noop {} {\bibfield
  {journal} {\bibinfo  {journal} {Computational Materials Science}\ }\textbf
  {\bibinfo {volume} {28}},\ \bibinfo {pages} {259 -- 273} (\bibinfo {year}
  {2003})}\BibitemShut {NoStop}%
\bibitem [{\citenamefont {Haule}\ \emph {et~al.}(2010)\citenamefont {Haule},
  \citenamefont {Yee},\ and\ \citenamefont {Kim}}]{Haule10supp}%
  \BibitemOpen
  \bibfield  {author} {\bibinfo {author} {\bibfnamefont {K.}~\bibnamefont
  {Haule}}, \bibinfo {author} {\bibfnamefont {C.-H.}\ \bibnamefont {Yee}}, \
  and\ \bibinfo {author} {\bibfnamefont {K.}~\bibnamefont {Kim}},\ }\bibfield
  {title} {\enquote {\bibinfo {title} {Dynamical mean-field theory within the
  full-potential methods: Electronic structure of $\text{CeIrIn}_{5}$,
  $\text{CeCoIn}_{5}$, and $\text{CeRhIn}_{5}$},}\ }\href {\doibase
  10.1103/PhysRevB.81.195107} {\bibfield  {journal} {\bibinfo  {journal} {Phys.
  Rev. B}\ }\textbf {\bibinfo {volume} {81}},\ \bibinfo {pages} {195107}
  (\bibinfo {year} {2010})}\BibitemShut {NoStop}%
\bibitem [{\citenamefont {Anisimov}\ \emph {et~al.}(1997)\citenamefont
  {Anisimov}, \citenamefont {Aryasetiawan},\ and\ \citenamefont
  {Lichtenstein}}]{LDA+Usupp}%
  \BibitemOpen
  \bibfield  {author} {\bibinfo {author} {\bibfnamefont {V.~I.}\ \bibnamefont
  {Anisimov}}, \bibinfo {author} {\bibfnamefont {F.}~\bibnamefont
  {Aryasetiawan}}, \ and\ \bibinfo {author} {\bibfnamefont {A.~I.}\
  \bibnamefont {Lichtenstein}},\ }\bibfield  {title} {\enquote {\bibinfo
  {title} {First-principles calculations of the electronic structure and
  spectra of strongly correlated systems: the $\text{LDA+U}$ method},}\
  }\href@noop {} {\bibfield  {journal} {\bibinfo  {journal} {J. Phys. Condens.
  Matter}\ }\textbf {\bibinfo {volume} {9}},\ \bibinfo {pages} {767} (\bibinfo
  {year} {1997})}\BibitemShut {NoStop}%
\bibitem [{\citenamefont {Jung}\ and\ \citenamefont
  {Gerber}(1996)}]{jung_vibrational_1996supp}%
  \BibitemOpen
  \bibfield  {author} {\bibinfo {author} {\bibfnamefont {Joon~O.}\ \bibnamefont
  {Jung}}\ and\ \bibinfo {author} {\bibfnamefont {R.~Benny}\ \bibnamefont
  {Gerber}},\ }\bibfield  {title} {\enquote {\bibinfo {title} {Vibrational wave
  functions and spectroscopy of ({H$_2$O})n, n=2,3,4,5: {Vibrational}
  self-consistent field with correlation corrections},}\ }\href {\doibase
  10.1063/1.472960} {\bibfield  {journal} {\bibinfo  {journal} {The Journal of
  Chemical Physics}\ }\textbf {\bibinfo {volume} {105}},\ \bibinfo {pages}
  {10332--10348} (\bibinfo {year} {1996})}\BibitemShut {NoStop}%
\bibitem [{\citenamefont {Stuart}\ \emph {et~al.}(1997)\citenamefont {Stuart},
  \citenamefont {Susan~J.},\ and\ \citenamefont {Joel~M.}}]{n-mode-1supp}%
  \BibitemOpen
  \bibfield  {author} {\bibinfo {author} {\bibfnamefont {Carter}\ \bibnamefont
  {Stuart}}, \bibinfo {author} {\bibfnamefont {Culik}\ \bibnamefont
  {Susan~J.}}, \ and\ \bibinfo {author} {\bibfnamefont {Bowman}\ \bibnamefont
  {Joel~M.}},\ }\bibfield  {title} {\enquote {\bibinfo {title} {Vibrational
  self-consistent field method for many-mode systems: A new approach and
  application to the vibrations of $\text{CO}$ adsorbed on $\text{Cu(100)}$},}\
  }\href@noop {} {\bibfield  {journal} {\bibinfo  {journal} {J. Chem. Phys.}\
  }\textbf {\bibinfo {volume} {107}},\ \bibinfo {pages} {10458} (\bibinfo
  {year} {1997})}\BibitemShut {NoStop}%
\bibitem [{\citenamefont {Emil~Lund}\ \emph {et~al.}(2018)\citenamefont
  {Emil~Lund}, \citenamefont {Bo}, \citenamefont {Ian~Heide},\ and\
  \citenamefont {Ove}}]{n-mode-3supp}%
  \BibitemOpen
  \bibfield  {author} {\bibinfo {author} {\bibfnamefont {Klinting}\
  \bibnamefont {Emil~Lund}}, \bibinfo {author} {\bibfnamefont {Thomsen}\
  \bibnamefont {Bo}}, \bibinfo {author} {\bibfnamefont {Godtliebsen}\
  \bibnamefont {Ian~Heide}}, \ and\ \bibinfo {author} {\bibfnamefont
  {Christiansen}\ \bibnamefont {Ove}},\ }\bibfield  {title} {\enquote {\bibinfo
  {title} {Employing general fit-bases for construction of potential energy
  surfaces with an adaptive density-guided approach},}\ }\href@noop {}
  {\bibfield  {journal} {\bibinfo  {journal} {J. Chem. Phys.}\ }\textbf
  {\bibinfo {volume} {148}},\ \bibinfo {pages} {064113} (\bibinfo {year}
  {2018})}\BibitemShut {NoStop}%
\bibitem [{\citenamefont {Carolin}\ and\ \citenamefont {Ove}(2016)}]{n-mode-4supp}%
  \BibitemOpen
  \bibfield  {author} {\bibinfo {author} {\bibfnamefont {K\"onig}\ \bibnamefont
  {Carolin}}\ and\ \bibinfo {author} {\bibfnamefont {Christiansen}\
  \bibnamefont {Ove}},\ }\bibfield  {title} {\enquote {\bibinfo {title}
  {Linear-scaling generation of potential energy surfaces using a double
  incremental expansion},}\ }\href@noop {} {\bibfield  {journal} {\bibinfo
  {journal} {J. Chem. Phys.}\ }\textbf {\bibinfo {volume} {145}},\ \bibinfo
  {pages} {064105} (\bibinfo {year} {2016})}\BibitemShut {NoStop}%
\bibitem [{\citenamefont {Stoll}(1992)}]{stoll_correlation_1992supp}%
  \BibitemOpen
  \bibfield  {author} {\bibinfo {author} {\bibfnamefont {Hermann}\ \bibnamefont
  {Stoll}},\ }\bibfield  {title} {\enquote {\bibinfo {title} {Correlation
  energy of diamond},}\ }\href {\doibase 10.1103/PhysRevB.46.6700} {\bibfield
  {journal} {\bibinfo  {journal} {Physical Review B}\ }\textbf {\bibinfo
  {volume} {46}},\ \bibinfo {pages} {6700--6704} (\bibinfo {year} {1992})},\
  \bibinfo {note} {publisher: American Physical Society}\BibitemShut {NoStop}%
\bibitem [{\citenamefont {Friedrich}\ and\ \citenamefont
  {Walczak}(2013)}]{friedrich_incremental_2013supp}%
  \BibitemOpen
  \bibfield  {author} {\bibinfo {author} {\bibfnamefont {Joachim}\ \bibnamefont
  {Friedrich}}\ and\ \bibinfo {author} {\bibfnamefont {Katarzyna}\ \bibnamefont
  {Walczak}},\ }\bibfield  {title} {\enquote {\bibinfo {title} {Incremental
  $\text{CCSD(T)(F12)}${\textbar}$\text{MP2-F12}$ --- $\text{A}$ method to
  obtain highly accurate $\text{CCSD(T)}$ energies for large molecules},}\
  }\href {\doibase 10.1021/ct300938w} {\bibfield  {journal} {\bibinfo
  {journal} {Journal of Chemical Theory and Computation}\ }\textbf {\bibinfo
  {volume} {9}},\ \bibinfo {pages} {408--417} (\bibinfo {year} {2013})},\
  \bibinfo {note} {publisher: American Chemical Society}\BibitemShut {NoStop}%
\bibitem [{\citenamefont {Zimmerman}(2017)}]{zimmerman_strong_2017supp}%
  \BibitemOpen
  \bibfield  {author} {\bibinfo {author} {\bibfnamefont {Paul~M.}\ \bibnamefont
  {Zimmerman}},\ }\bibfield  {title} {\enquote {\bibinfo {title} {Strong
  correlation in incremental full configuration interaction},}\ }\href
  {\doibase 10.1063/1.4985566} {\bibfield  {journal} {\bibinfo  {journal} {The
  Journal of Chemical Physics}\ }\textbf {\bibinfo {volume} {146}},\ \bibinfo
  {pages} {224104} (\bibinfo {year} {2017})}\BibitemShut {NoStop}%
\bibitem [{\citenamefont {Stoll}(2019)}]{stoll_toward_2019supp}%
  \BibitemOpen
  \bibfield  {author} {\bibinfo {author} {\bibfnamefont {Hermann}\ \bibnamefont
  {Stoll}},\ }\bibfield  {title} {\enquote {\bibinfo {title} {Toward a
  wavefunction-based treatment of strong electron correlation in extended
  systems by means of incremental methods},}\ }\href@noop {} {\bibfield
  {journal} {\bibinfo  {journal} {The Journal of Chemical Physics}\ }\textbf
  {\bibinfo {volume} {151}},\ \bibinfo {pages} {044104} (\bibinfo {year}
  {2019})},\ \bibinfo {note} {publisher: American Institute of
  Physics}\BibitemShut {NoStop}%
\bibitem [{\citenamefont {Sobol}(2001)}]{sobol_global_2001supp}%
  \BibitemOpen
  \bibfield  {author} {\bibinfo {author} {\bibfnamefont {I.~M}\ \bibnamefont
  {Sobol}},\ }\bibfield  {title} {{\selectlanguage {english}\enquote {\bibinfo
  {title} {Global sensitivity indices for nonlinear mathematical models and
  their $\text{M}$onte $\text{C}$arlo estimates},}\ }}\href@noop {} {\bibfield
  {journal} {\bibinfo  {journal} {Mathematics and Computers in Simulation}\
  }\bibinfo {series} {The {Second} {IMACS} {Seminar} on {Monte} {Carlo}
  {Methods}},\ \textbf {\bibinfo {volume} {55}},\ \bibinfo {pages} {271--280}
  (\bibinfo {year} {2001})}\BibitemShut {NoStop}%
\bibitem [{\citenamefont {Griebel}(2006)}]{Griebel:2005supp}%
  \BibitemOpen
  \bibfield  {author} {\bibinfo {author} {\bibfnamefont {M.}~\bibnamefont
  {Griebel}},\ }\bibfield  {title} {\enquote {\bibinfo {title} {Sparse grids
  and related approximation schemes for higher dimensional problems},}\ }in\
  \href@noop {} {\emph {\bibinfo {booktitle} {Foundations of Computational
  Mathematics (FoCM05), Santander}}},\ \bibinfo {editor} {edited by\ \bibinfo
  {editor} {\bibfnamefont {L.}~\bibnamefont {Pardo}}, \bibinfo {editor}
  {\bibfnamefont {A.}~\bibnamefont {Pinkus}}, \bibinfo {editor} {\bibfnamefont
  {E.}~\bibnamefont {Suli}}, \ and\ \bibinfo {editor} {\bibfnamefont {M.J.}\
  \bibnamefont {Todd}}}\ (\bibinfo  {publisher} {Cambridge University Press},\
  \bibinfo {year} {2006})\ pp.\ \bibinfo {pages} {106--161}\BibitemShut
  {NoStop}%
\bibitem [{\citenamefont {Chen}\ \emph {et~al.}(2019)\citenamefont {Chen},
  \citenamefont {Wang}, \citenamefont {Ye},\ and\ \citenamefont
  {Hu}}]{chen_comparative_2019supp}%
  \BibitemOpen
  \bibfield  {author} {\bibinfo {author} {\bibfnamefont {Liming}\ \bibnamefont
  {Chen}}, \bibinfo {author} {\bibfnamefont {Hu}~\bibnamefont {Wang}}, \bibinfo
  {author} {\bibfnamefont {Fan}\ \bibnamefont {Ye}}, \ and\ \bibinfo {author}
  {\bibfnamefont {Wei}\ \bibnamefont {Hu}},\ }\bibfield  {title}
  {{\selectlanguage {english}\enquote {\bibinfo {title} {Comparative study of
  {HDMRs} and other popular metamodeling techniques for high dimensional
  problems},}\ }}\href@noop {} {\bibfield  {journal} {\bibinfo  {journal}
  {Structural and Multidisciplinary Optimization}\ }\textbf {\bibinfo {volume}
  {59}},\ \bibinfo {pages} {21--42} (\bibinfo {year} {2019})}\BibitemShut
  {NoStop}%
\end{thebibliography}


\clearpage

%\pagebreak and \newpage did not work here

%%%%%%%%%%%Supplement%%%%%%%%%%%%%%%%%%%
%%%%%%%%%% Merge with supplemental materials %%%%%%%%%%

\title{Bypassing the computational bottleneck of quantum-embedding theories for \\ strong electron correlations with machine learning}

\author{John Rogers}
\affiliation{Department of Physics and Astronomy, Texas A\&M University, College Station, Texas 77845, USA}
\author{Tsung-Han Lee}
\affiliation{Physics and Astronomy Department, Rutgers University, Piscataway, New Jersey 08854, USA}
\author{Sahar Pakdel}
\affiliation{Department of Physics and Astronomy, Aarhus University, 8000,
Aarhus C, Denmark}
\author{Wenhu Xu}
\affiliation{Condensed Matter Physics and Materials Science Department, Brookhaven National Laboratory, Upton, NY 11973}
\author{Vladimir Dobrosavljevi\'c}
\affiliation{Department of Physics and National High Magnetic Field Laboratory, Florida State University, Tallahassee, Florida 32306, USA}
\author{Yong-Xin Yao}
%\altaffiliation{Corresponding author: ykent@iastate.edu}
\affiliation{Ames Laboratory-U.S. DOE and Department of Physics and Astronomy, Iowa State University, Ames, Iowa 50011, USA}
\author{Ove Christiansen}
\altaffiliation{Corresponding author: ove@chem.au.dk}
\affiliation{Department of Chemistry, Aarhus University, 8000, Aarhus C, Denmark}
\author{Nicola Lanat\`a}
\altaffiliation{Corresponding author: lanata@phys.au.dk}
\affiliation{Department of Physics and Astronomy, Aarhus University, 8000, Aarhus C, Denmark}

\date{\today} 

\maketitle

\widetext
\begin{center}
\textbf{\large Supplemental material for: Bypassing the computational bottleneck of quantum-embedding theories for strong electron correlations with machine learning}
\end{center}

%%%%%%%%%% Prefix a "S" to all equations, figures, tables and reset the counter %%%%%%%%%%
\setcounter{equation}{0}
\setcounter{figure}{0}
\setcounter{section}{0}
\setcounter{table}{0}
\setcounter{page}{1}
\makeatletter

%remove these to get rid of S
%\renewcommand{\theequation}{S\arabic{equation}}
%\renewcommand{\thefigure}{S\arabic{figure}}
%\renewcommand{\bibnumfmt}[1]{[S#1]}
%\renewcommand{\citenumfont}[1]{S#1}\
%\renewcommand{\thesection}{S\arabic{section}}

\author{John Rogers}
\affiliation{Department of Physics and Astronomy, Texas A\&M University, College Station, Texas 77845, USA}
\author{Tsung-Han Lee}
\affiliation{Physics and Astronomy Department, Rutgers University, Piscataway, New Jersey 08854, USA}
\author{Sahar Pakdel}
\affiliation{Department of Physics and Astronomy, Aarhus University, 8000,
Aarhus C, Denmark}
\author{Wenhu Xu}
\affiliation{Condensed Matter Physics and Materials Science Department, Brookhaven National Laboratory, Upton, NY 11973}
\author{Vladimir Dobrosavljevi\'c}
\affiliation{Department of Physics and National High Magnetic Field Laboratory, Florida State University, Tallahassee, Florida 32306, USA}
\author{Yong-Xin Yao}
%\altaffiliation{Corresponding author: ykent@iastate.edu}
\affiliation{Ames Laboratory-U.S. DOE and Department of Physics and Astronomy, Iowa State University, Ames, Iowa 50011, USA}
\author{Ove Christiansen}
\altaffiliation{Corresponding author: ove@chem.au.dk}
\affiliation{Department of Chemistry, Aarhus University, 8000, Aarhus C, Denmark}
\author{Nicola Lanat\`a}
\altaffiliation{Corresponding author: lanata@phys.au.dk}
\affiliation{Department of Physics and Astronomy, Aarhus University, 8000, Aarhus C, Denmark}

\date{\today}

%%\title{Supplemental material for: Combining quantum-embedding theories with machine learning}

%%%\title{Supplemental material for: Bypassing the computational bottleneck of quantum-embedding theories for strong electron correlations with machine learning}

% 
%%\pacs{64, 71.30.+h, 71.27.+a}
% 64 (Equations of state, phase equilibria, and phase transitions)
% 71.30.+h (Metal-insulator transitions and other electronic transitions)
% 71.27.+a (Strongly correlated electron systems; heavy fermions)

\section{The GA Method}\label{ga}

For completeness, here we briefly summarize the equations underlying the formulation of the
GA as a QE scheme,
which was previously derived in Refs.~\cite{Our-PRXsupp,Lanata2016supp}.

The GA solution is obtained by calculating the saddle-points
of the following Lagrange function~\cite{Lanata2016supp}:

%\begin{widetext}
\bea
&&\Lag_N[U,J,E;\, \Phi,E^c,  \R,\lambda,\mu, \D, \lambda^{c},\Delta]
=\nonumber\\
&&\qquad\quad
\frac{1}{\mathcal{N}}\left[\Av{\Psi_0}{\h_{\text{qp}}[\R,\lambda;\mu]}
+E\!\left(1\!-\!\langle\Psi_0|\Psi_0\rangle\right)\right]
+ \sum_i\left[\Av{\Phi_i}{\h_i^{\text{emb}}[\D_i,\lambda_i^c]}
+ E^c_i\!\left(1-\langle \Phi_i | \Phi_i \rangle
\right)\right]-
\nonumber\\
&&\qquad\quad
-\sum_i\bigg[
\sum_{ab=1}^{M_i}\big(
\left[\lambda_i\right]_{ab}+\left[\lambda^c_i\right]_{ab}\big)\left[\Delta_{i}\right]_{ab}
+\sum_{c a \alpha=1}^{M_i}\left(
\left[\D_{i}\right]_{a\alpha}\left[\R_{i}\right]_{c\alpha}
\big[\Delta_{i}(1-\Delta_{i})\big]^{\frac{1}{2}}_{c\alpha}
+\text{c.c.}\right)
\bigg] + \mu\,N
\,,
\label{Lag-Psi0-Phi}
\eea
%\end{widetext}

where:
\begin{align}
\h_{\text{qp}}&=
\sum_{\bk}\sum_{ij=1}^{\eta}
\sum_{a=1}^{M_i}\sum_{b=1}^{M_j}
\big[\R^\dagga_i {\epsilon}_{\bk,ij}
\R_j^\dagger+\lambda_i-\mu\big]_{ab}\,\fc_{\bk i a}\fa_{\bk i b}
\label{hqp}
\\
\h_i^{\text{emb}}&=
\hat{\mathcal{H}}^{\text{loc}}_i[U_i,J_i,E_i]
%%[\{\hat{c}^\dagger_{i \alpha}\},\{\hat{c}^\dagga_{i \alpha}\}] 
+
%\nonumber\\ +&
\sum_{a \alpha=1}^{M_i} \left(
\left[\D_{i}\right]_{a\alpha}
\hat{c}^\dagger_{i \alpha}\hat{f}^\dagga_{i a}+\text{H.c.}\right)
+\sum_{a b=1}^{M_i} \left[\lambda^c_{i}\right]_{ab}
\hat{f}^\dagga_{i b}\hat{f}^\dagger_{i a} 
%\label{h-emb-i}
\,,
\label{hemb}
\end{align}
$\R,\D,\lambda,\lambda^c,\Delta$
are complex block-matrices whose respective $M_i\times M_i$ 
blocks are
$\R_i,\D_i,\lambda_i,\lambda_i^c,\Delta_i$, where
$\R_i,\lambda_i,\lambda^c_i$ are hermitian,
$\mu$ is the chemical potential,
$N$ is the total number of electrons in the system
(normalized to the number of $k$-points $\mathcal{N}$),
$\epsilon_{\bk,ij}$ are matrices constituted by $M_i\times M_j$
blocks labeled by $i,j$ with entries $\epsilon^{\alpha\beta}_{\bk,ij}$ and $\ket{\Psi_0}$
is the most general single-particle wavefunction within the space of $\h_{\text{qp}}$,
see Eq.~\eqref{hqp}.
By construction~\cite{Our-PRXsupp}, the "embedding states"
$\ket{\Phi_i}$ are assumed to lie within the $M_i$-particle
subspace of $\h_i^{\text{emb}}$, see Eq.~\eqref{hemb}, i.e., they satisfy the following equation:
\be
\big[\sum_{a}\hat{c}^\dagger_{i \alpha}\hat{c}^\dagga_{i \alpha}
+\sum_a \hat{f}^\dagger_{ia}\hat{f}^\dagga_{ia}\big]\ket{\Phi_i}
=M_i\,\ket{\Phi_i}\,.
\label{half-filling}
\ee

Physical observables can be calculated from the parameters of the theory 
realizing the saddle-point of Eq.~\eqref{Lag-Psi0-Phi}.
In particular, the total energy of the system equals the saddle-point value of $\Lag_N$.
The expectation values of any local operator $\hat{O}\big[\{\cc_{\bR i\alpha},\ca_{\bR i\alpha}\}\big]$ can be calculated as follows~\cite{Our-PRXsupp}:
\begin{equation}
\langle \hat{O}\big[\{\cc_{\bR i\alpha},\ca_{\bR i\alpha}\}\big] \rangle=\langle\Phi_i| \hat{O}\big[\{\hat{c}_{i\alpha}^\dagger,\hat{c}_{i\alpha}^\dagga\}\big] |\Phi_i\rangle\,, \label{eq:local_observables}
\end{equation} 
where $|\Phi_i\rangle$ is the ground state of the 
self-consistent EH 
$\h_i^{\text{emb}}[U_i,J_i,E_i,\D_i,\lambda_i^c]$.
The local self energy, instead, is expressed in terms of the
self-consistent parameters $\R_i$ and
$\lambda_i$ as follows~\cite{Gebhard-FLsupp}:
\begin{equation}
\Sigma_i(\omega)=-(\omega+\mu) \big[1-\R_i^\dagger \R_i^\dagga\big]\big[\R_i^\dagger \R_i^\dagga\big]^{-1}+\R_i^{-1}\lambda_i \R_i^{\dagger -1}
\,.\label{SE_RISB}
\end{equation}

The most expensive operation necessary for obtaining the saddle-point of Eq.~\eqref{Lag-Psi0-Phi}
is to compute recursively the ground-state of the EH, see Eq.~\eqref{hemb}. 
The nKRR methodology proposed in this work allows us to bypass altogether this operation.

\subsection*{GA+DFT}

Within DFT+GA, the Kohn-Sham parameters ${\epsilon}_{\bk,ij}$ and $E$ are updated at each charge iteration and determined self-consistently.
Evaluating these parameters and calculating the electron density at each charge iteration is essentially as expensive as in all classic DFT implementations.

Our LDA and LDA+GA calculations were performed utilizing the DFT code WIEN2k~\cite{WIEN2ksupp}.
The LDA+GA solver was implemented following Ref.~\cite{Lanata2016supp}.
The LAPW interface between WIEN2k and the RISB was implemented as described in Ref.~\cite{Haule10supp}, utilizing the fully-localized limit (FFL) double-counting functional~\cite{LDA+Usupp}.
All calculations were performed setting $RKmax=9$. The convergence with respect to the number of $k$-points was verified for all systems considered.

\section{The $n$-mode representation}\label{3d} % (may go into appendix)}

As mentioned in in the main text, applying directly KRR for learning a multivariate function
$F(X_1,X_2,\dots,X_d)$ from data built on a
mesh with $m$ data points per axis requires $N\sim m^d$ function evaluations.
Therefore, the complexity of the learning problem grows exponentially as a function of the number of variables $d$.

The basic idea underlying the $n$-mode representation is to represent a high-dimensional function $F$
in terms of the following lower-dimensional functions:
%The $n$-mode representation is an incremental expansion method for approximating multivariate functions, that will allow us to reduce considerably the number of data required.
%Following Ref.~X, we define the series of $d$ functions:
%\begin{widetext}
\begin{align}
%&\bar{F}^0D\equiv F^0\\
\bar{F}^{1}_i(X_i) &= F^{1}_i(X_i)-F^0
\label{fbar1}
\\
\bar{F}^{2}_{ij}(X_i,X_j) &= F^{2}_{ij}(X_i,X_j)-\bar{F}^{1}_i(X_i)
%\\&\hspace{100}
-\bar{F}^{1}_j(X_j)-F^0 \nonumber \\
\bar{F}^{3}_{ijk}(X_i,X_j,X_k)&=
F^{3}_{ijk}(X_i,X_j,X_k)
%\\&\hspace{10}
-\bar{F}^{2}_{ij}(X_i,X_j) - \bar{F}^{2}_{ik}(X_i,X_k) - \bar{F}^{2}_{jk}(X_j,X_k) 
%\nonumber \\
%&\hspace{25}
-\bar{F}^{1}_i(X_i)-\bar{F}^{1}_j(X_j)-\bar{F}^{1}_k(X_k)-F^0 \nonumber
\,,
\end{align}
%\end{widetext}
etc.., where:
\begin{align}
F^0 &=  F(0,0,\dots,0,0,0,\dots,0,0,0,\dots,0)
\nonumber\\ 
F^{1}_i(X_i) &=  F(0,0,\dots,0,X_i,0,\dots,0,0,0,\dots,0)
\nonumber\\ 
F^{2}_{ij}(X_i,X_j) &=  F(0,0,\dots,0,X_i,0,\dots,0,X_j,0\dots,0)
\,,
\label{cutsupp}
\end{align}
etc..,
are the so-called ``cut-functions,'' which
are restrictions of $F(X)$ to hyperplanes where
subsets of the components of $X$ are set to $0$.
Specifically, the $n$-mode representation of $F(X)$ is given by the following equation:
\be
   F(X_1,X_2,\dots,X_d)=F^0+\sum_{i=1}^{d}\bar{F}^{1}_i(X_i)
   +\sum_{j>i=1}^{d}\bar{F}^{2}_{ij}(X_i,X_j)+\sum_{k>j>i=1}^d\bar{F}^{3}_{ijk}(X_i,X_j,X_k)+...
   \label{fbarcollect}
\ee

It can be readily verified that, when all terms are retained,
Eq.~\eqref{fbarcollect} is an exact identity (which is a major advantage compared to other approximations such as the Taylor expansion).
Truncating 
this series up to a given order $n<d$ provides us with an approximation that is exact only over the domains of the order-$n$ cut-functions, while elsewhere it is an approximation that tends to be more accurate in the proximity of the domains of the order-$n$ cut functions.

By inspecting Eq.~\eqref{fbarcollect} we note that, 
building a mesh of training data with $m$ points per axis, the nKRR method up to order $n$ requires only the following number of data points:
\be
N^{(n)}={n\choose d} \cdot {m^n}\,,
\label{N_nm}
\ee
which scales polynomially (as $d^n$) as a function of $d$,
rather than exponentially.
%where the exponent of $m$ is only $n<d$.

Note that, as opposed to other dimensionality-reduction methods ---where the number of effective input variables is decreased,--- the $n$-mode expansion includes from the outset \emph{all} variables.

\subsection*{Variable shifts}

Whenever it is possible to estimate the range of input values where $F$ has to be evaluated for a particular application (e.g., based on physical arguments inherent in the particular context of application), performing a suitable change of variables 
$X=v(Y)$
can reduce significantly the necessary truncation order $n$.
In fact, a change of variables can be designed in such a way that the relevant range of input values is as close as possible to the  domains of the cut functions of
$G(Y)=F(v(Y))$,
where the $n$-mode expansion is more accurate (by construction).

In particular, in the main text we have exploited this freedom to shift the origin with a change of variables of the form:
\be
Y=X-\bar{X}\,, 
\ee
which improved considerably the speed of convergence of the $n$-mode expansion, facilitating the learning problem.
The numerical values of the components of $\bar{X}_1$ and $\bar{X}_2$ are reported 
in Table~\ref{table:z}, while $\bar{X}_3$, $\bar{X}_4$, $\bar{X}_5$ have been all set to $-0.694$\,eV (based on a single DFT+GA calculation of $\delta$-Pu at its experimental equilibrium volume).

\begin {table}[ht]
\caption{Pre-calculated values of the components of $\bar{X}$ (eV)}\label{table:z}
\smallskip
\centering
\begin{tabular}{| c | c | c | c | c | c |}
\hline
\textbf{} & \textbf{Pa} & \textbf{U} & \textbf{Np}  & \textbf{Pu}  & \textbf{Am}\\
\hline
$\bar{X}_1$          & -1.497      & -4.218     & -6.939   & -9.660   & -12.381  \\
\hline
$\bar{X}_2$          & -0.422  & -0.422    &  -0.490   & -0.558  & -0.558  \\
%\hline
%$\bar{X}_3$          & \multicolumn{5}{c|}{-0.051} \\
%\hline
%$\bar{X}_4$          & \multicolumn{5}{c|}{-0.051} \\
%\hline
%$\bar{X}_5$          & \multicolumn{5}{c|}{-0.051} \\
\hline
\end{tabular}
\end{table}

\subsection*{Previous applications of the $n$-mode representation}

While in this work we have applied the $n$-mode representation within the context of quantum embedding methods, this method has been originally designed and explored in different contexts (and under different names).
In particular, as mentioned in the main text, the $n$-mode representation has recently gained significant attention for representing and computing potential energy surfaces (PES) of molecules~\cite{jung_vibrational_1996supp,n-mode-1supp,n-mode-3supp,n-mode-4supp}.

Also the so-called "incremental method"~\cite{stoll_correlation_1992supp} and "many-body expansion" for computing electronic energies of molecules can be considered as a variations to the $n$-mode expansions~\cite{n-mode-4supp}. 
These tools have both been exploited for obtaining and representing accurate electronic energies of large molecules and molecular clusters~\cite{friedrich_incremental_2013supp}.
Furthermore, the incremental method has also been applied previously with a focus on strong electron correlation~\cite{zimmerman_strong_2017supp,stoll_toward_2019supp}.
Specifically, these works applied the incremental expansion directly to the solution of the Schr{\"o}dinger equation for solving specific chemical problems.
The incremental methods for single point electronic energies
has been also combined with the PES n-mode expansion to obtain a double incremental expansion of the PES paving the way for obtaining linear scaling construction of PESs ---an otherwise hard-to-imagine result~\cite{n-mode-4supp}.
Finally, the $n$-mode representation can be seen as one variant of high-dimensional model representation (HDMR) and is sometimes denoted cut-HDMR. In turn, HDMR is closely related to the ANOVA method of statistics~\cite{sobol_global_2001supp},
and it has been applied, from this side, for sparse-grid methods in high-dimensional problems~\cite{Griebel:2005supp}.
Recently  cut-HDMR is receiving significant attention in other fields, for example machine learning in engineering~\cite{chen_comparative_2019supp}.

\section{Implementation of KRR method}\label{krr}

Kernel ridge regression (KRR) is a non-parametric form of  regression. Here we describe the specific procedure utilized in the calculations presented in this work.

Given a continuous function $F(X)$, the KRR method provides us with an approximation represented as follows:
\be
\tilde{F}_{\sigma}(X)=\sum_{l=1}^{N} \alpha_l\, k_{\sigma}(X_l,X)\,,
\label{ftilde}
\ee
where $X\in\mathbb{R}^d$, $\{X_l\in\mathbb{R}^d \;|\;  l=1,\dots,N\}$ is a set of points belonging to the domain of $F$ (known as "feature vectors"), and $k$ is the so-called "kernel" function.
Specifically, in this this work we utilized the so-called the "radial basis function (RBF)" kernel (also known as the Gaussian kernel), which is defined as follows:
\be
k_\sigma(A,B)=\text{exp}\left(-\frac{\|A-B\|^2}{2\sigma^2}\right)\,,
\label{rbf}
\ee
where $\|X\|^2=\sum_{m=1}^N |X_m|^2$ is the standard Euclidean norm.

The procedure for determining the coefficients $\alpha_l$
and the kernel width parameter $\sigma$ is the following:
\begin{itemize}
\item A "training data" set $\{Y_l=F(X_l)\in\mathbb{R}^d \;|\;  l=1,\dots,N\}$ is constructed by evaluating $F$ on the feature vectors $\{X_l\in\mathbb{R}^d \;|\;  l=1,\dots,N\}$.
\item The following minimization is performed:
\begin{align}
    \bar{\alpha}(\sigma,\lambda)&=
    \argmin_{\alpha' \in \mathbb{R}^d}
    \left[\sum_{m=1}^N 
    \left(
    \sum_{l=1}^{N} \alpha'_l(\sigma)\, k_{\sigma}(X_l,X_m)
    -Y_m\right)^2
    +\lambda \sum_{l,m=1}^{N} 
    \alpha'_l\, k_\sigma(X_l,X_m)\, \alpha'_m
     \right]
     \label{cost}
     \\
     &=[K_\sigma+\lambda I]^{-1}Y\,,
     \label{matalpha}
\end{align}
where $[K_\sigma]_{lm}=k_\sigma(X_l,X_m)$ is the kernel matrix,
$[Y]_m=Y_m$ is the training-data-set vector and the right member of Eq.~\eqref{cost} is called the "cost function".

\item The coefficients $\alpha_l$ of Eq.~\eqref{ftilde} are:
\be
\alpha_l=[\bar{\alpha}(\sigma,\lambda)]_l\,,
\ee
where $\sigma$ and $\lambda$, typically named "hyperparameters," are determined using the so-called "cross-validation" (CV) method, that is an empirical protocol designed to optimize the predictive power of the KRR model.
In particular, $\lambda$ is a regularization parameters that can be used to remove singularities in Eq.~\eqref{matalpha} and avoiding over-fitting.

\end{itemize}

The standard k-fold procedure consists in dividing the data set into a finite number $b$ of batches (folds) with similar size.
The ML solver is trained using $b-1$ of these batches at a given pair of $\lambda$ and $\sigma$ values. The (trained) ML solver is subsequently used for predicting the data points in the excluded fold, and the errors in the predictions are measured. This procedure is repeated, excluding each fold once and keeping a running total of the 
error for the given pair of hyperparameters. A mesh of possible combinations of $\lambda$ and $\sigma$ are tested as described above, and the pair that yields the smallest error is chosen.

In our work we employed a variation to the classic k-fold CV outlined above. Specifically, we have restricted the optimization of the hyperparameters $\lambda$ and $\sigma$ to values such that $\|\lambda\|\geqslant 10^{-6}$.
The purpose of this cutoff was to reduce spurious high-frequency components in the KRR function approximation [Eq.~\eqref{ftilde}]. This improved the overall stability of the GA+nKRR algorithm, without compromising the accuracy of the nKRR solver.

\subsection*{Data-set mesh}

By construction, the meshes of the EH input variables $Y_i$
---utilized in our calculations for training the KRR solver---
are symmetric around 0.
The respective number of data points per axis $m_i$, as well as the
 minimum $Y_i^{\text{min}}$ and maximum $Y_i^{\text{max}}$ values of the corresponding intervals,  are reported in Table~\ref{table:mesh}.

\begin {table}[ht]
\caption{Parameters of the training data mesh for the components $Y_i$ of $Y$}\label{table:mesh}
\smallskip
\centering
\begin{tabular}{| c | c | c | c | c | c |}
\hline
~ & $Y_i^{\text{min}}$ (eV) & $Y_i^{\text{max}}$ (eV) & $m_i$ \\
\hline
$Y_1$          & -0.340      &  0.340     & 9    \\
\hline
$Y_2$          & -0.068  &  0.068    & 5     \\
\hline
$Y_3$          & -0.136 & 0.136  & 17      \\
%\hline
$Y_4$          & -0.136 & 0.136  & 17    \\
%\hline
$Y_5$          & -0.136 & 0.136  & 17      \\
\hline
\end{tabular}
\end{table}

\end{document}